\documentclass[a4paper,11pt]{article}

\usepackage{amsmath, amssymb, a4wide, bm, graphicx, subfig, cancel, youngtab}
\usepackage{cite}
\usepackage{hyperref}

\newcommand{\SU}[1]{\mathrm{SU}(#1)}

\newcommand{\U}[1]{\mathrm{U}(#1)}

\newcommand{\nbrack}[1]{\left(#1\right)}
\newcommand{\sbrack}[1]{\left[#1\right]}

\newcommand{\expect}[1]{\langle#1\rangle}
\newcommand{\sep}[1]{\quad\mbox{#1} \quad}
\newcommand{\brm}[1]{\bm{\mathrm{#1}}}
\newcommand{\cbrm}[1]{\overline{\bm{\mathrm{#1}}}}

\def\asymm{\tiny\Yvcentermath1\yng(1,1)}
\def\fund{\tiny\Yvcentermath1\yng(1)}
\def\afund{\tiny\tilde{\Yvcentermath1\yng(1)}}
\def\widerow{\rule{0pt}{2.5ex}\rule[-1.5ex]{0pt}{0pt}}
\def\be{\begin{equation}}
\def\ee{\end{equation}}
\def\ba{\begin{eqnarray}}
\def\ea{\end{eqnarray}}
\def\cN{{\cal N}}
\def\uno{\mbox{1 \kern-.59em {\rm l}}}

\numberwithin{equation}{section}
\numberwithin{figure}{section}
\numberwithin{table}{section}

\begin{document}

\title{{\normalsize IPPP/10/33 DCPT/10/66\hfill\mbox{}\hfill\mbox{}}\\
\vspace{2.5 cm}
\Large{\textbf{Strong coupling, discrete symmetry and flavour}}}
\vspace{2.5 cm}
\author{Steven Abel and James Barnard\\[3ex]
\small{\em Institute for Particle Physics Phenomenology and Department of Mathematical Sciences,}\\
\small{\em Durham University, Durham DH1 3LE, UK}\\[1.5ex] 
\small s.a.abel or james.barnard@durham.ac.uk\\[1.5ex]}
\date{}
\maketitle
\vspace{2ex}
\begin{abstract}
\noindent We show how two principles -- strong coupling and discrete symmetry -- can work together to generate the flavour structure of the Standard Model.  We propose that in the UV the \emph{full} theory has a discrete flavour symmetry, typically only associated with tribimaximal mixing in the neutrino sector.  Hierarchies in the particle masses and mixing matrices then emerge from multiple strongly coupled sectors that break this symmetry.  This allows for a realistic flavour structure, even in models built around an underlying grand unified theory.  We use two different techniques to understand the strongly coupled physics: confinement in $\cN=1$ supersymmetry and the AdS/CFT correspondence.  Both approaches yield equivalent results and can be represented in a clear, graphical way where the flavour symmetry is realised geometrically.
\end{abstract}

\section{Introduction}

It is widely believed that strong coupling could explain several difficult phenomenological puzzles. The first truly compelling indication came with its application to supersymmetry breaking in ${\cal N}=1$ theories \cite{Affleck:1983rr, Affleck:1983mk, Affleck:1984xz}.  Subsequently the advent of Seiberg duality\cite{Seiberg:1994pq, Intriligator:1995au} opened rigorous lines of attack on other questions and led to a burst of model building activity \cite{Strassler:1995ia, Nelson:1996km, Berkooz:1997bb, ArkaniHamed:1997fq, Luty:1998vr}. Our purpose in this paper is to revisit one of those early ideas, namely the proposal of Refs.\cite{Strassler:1995ia,Nelson:2000sn} that strong coupling effects generate the hierarchical flavour structure that we observe in the quark and lepton masses and mixings\footnote{Although the idea that part of the Standard Model was composite was certainly not new at the time \cite{Volkas:1987cd}, Ref.\cite{Strassler:1995ia} is, to our knowledge, the first direct applications of Seiberg duality to it.}.

Developments since that work are encouraging a return to the subject \cite{Franco:2009wf, Poland:2009yb, Craig:2009hf, delAguila:2010vg, Kadosh:2010rm, Craig:2010ip, SchaferNameki:2010iz}. Firstly there is now a better understanding of the neutrino masses and mixings and a large number of well defined flavour models for the lepton sector (see Ref.\cite{Altarelli:2010gt} for a recent comprehensive review).  Mostly these rely on non-abelian discrete symmetries, the default example being the $A_{4}$ tetrahedral symmetry \cite{Babu:2002dz, Altarelli:2006kg, Csaki:2008qq, A4refs}.  Success has also been had with symmetry groups $T^{\prime}$ \cite{TPrefs}, $S_{4}$ \cite{S4refs}, $\Delta(27)$ \cite{D27refs} and other, more exotic groups \cite{ODSrefs}.  Strangely though, such discrete symmetries are not particularly helpful in explaining the quark sector masses and mixings. In order to accommodate the very different structures in the quark and lepton couplings, one has to have peculiar vacuum misalignments. Not only are these difficult to achieve, but also they do not sit easily with an underlying grand unified theory (GUT) structure.  Some progress has, however, been made in this direction: see Ref.~\cite{Kadosh:2010rm, GUTrefs} for some example models.

The second reason we wish to revisit the subject is that, in the interim, the AdS/CFT correspondence has given us a geometrical way to think about strong coupling.  Interest in strong coupling in flavour physics has usually been drawn to models with a single confinement scale.  The hierarchies either being associated with different dimensionalities of the composite particles \cite{Kaplan:1997tu, Haba:1997bj, Haba:1998wf, Franco:2009wf, Craig:2009hf, Craig:2010ip, SchaferNameki:2010iz}, exponentially small wavefunction overlaps in the extra dimension (which from the AdS/CFT point of view is equivalent) \cite{ArkaniHamed:1999dc, ArkaniHamed:1999za, Mirabelli:1999ks, Grossman:1999ra, Gherghetta:2000qt, Huber:2000ie, Gabella:2007cp, Csaki:2008qq, delAguila:2010vg, Kadosh:2010rm}, residual horizontal symmetries \cite{Luty:1996jd} or some combination of the three. However, the role of discrete symmetries such as $A_{4}$ in any of these scenarios is obscure: although it can be incorporated into the model \cite{Csaki:2008qq, Kadosh:2010rm} the underlying geometric origin is elusive\footnote{We should add that geometrical explanations of discrete symmetries in perturbative higher dimensional orbifold models can be achieved by compactifying on an orbifold that has the symmetry in question \cite{Watari:2002fd, Watari:2002tf, Altarelli:2006kg, Adulpravitchai:2009id, Burrows:2009pi}. String theory compactifications have been found that exploit similar geometric pictures either using orbifolds directly or D-brane configurations\cite{Kobayashi:2004ya, Kobayashi:2006wq, Ko:2007dz, Abe:2009vi, Abe:2010ii} with magnetized D-branes and wave-function localization.}.

\begin{figure}[!t]
\begin{center}
\includegraphics[width=6.5cm]{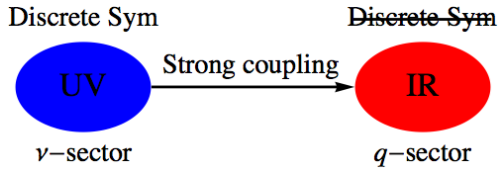}
\caption{A schematic diagram of our aim.  The UV theory will be bestowed with a discrete symmetry that is broken in the IR by strong coupling effects.  The neutrinos are taken to live in the UV as elementary fields so do not feel this symmetry breaking, whereas the quarks are taken to be composite operators living mainly in the IR and so do.\label{fig:schematic}}
\end{center}
\end{figure}

In this paper our approach will, by contrast, be governed by two principles. First we will assume that strong coupling effects (for example vastly different compositeness scales) are responsible for generating hierarchies in masses and mixings.  Second we believe that any discrete symmetries underlying the lepton/neutrino masses and mixings are also present in the quark sector but are broken by these strong coupling effects (in a way to be described). This is hard to avoid if one is interested in grand unified theories -- indeed one of the central points we wish to make is that strong coupling allows one to get much closer to an underlying GUT structure than purely perturbative models of flavour.  The general picture we will explore (depicted in Figure \ref{fig:schematic}) is related to that of Ref.\cite{Nelson:2000sn}, except that instead of anarchy in the UV there is discrete symmetry. The quark sector is hierarchical because quarks are mainly composite states whereas the leptons, and especially the neutrinos, are more elementary so display a discrete symmetry. We should stress at this point that the compositeness scales we have in mind could be anywhere between the weak and GUT scales.  In the latter case  we will invoke supersymmetry to protect the Higgs mass terms.

There are various ways to think about strong coupling. In the context of $\mathcal{N}=1$ supersymmetric theories Seiberg duality is a powerful tool because one has detailed information about the superpotential \cite{Seiberg:1994pq}. We begin by discussing strong coupling effects in that framework. In the models we will be examining first we introduce three strongly coupled $\SU2$ gauge groups, one for each generation. Each of these groups confines and produces part of a generation of Standard Model matter as bound states\footnote{Multiple confinement models similar to those we will be considering have appeared in the context of D-brane constructions in type IIA string theory \cite{Kitazawa:2004hz, Kitazawa:2004nf}. In those cases the compositeness scales were taken to be non-hierarchichal and, with the emphasis being on the localization of open strings at different brane intersections, these models can be thought of as a dual version of the localized bulk zero modes.} (in terms of the parent $\SU5$ theory of the Standard Model these states correspond to the antisymmetric \textbf{10}). The confining groups all have their own dynamics, naturally allowing an exponential hierarchy in the compositeness scales. This hierarchy appears in the superpotential  when one trades bound states of the electric theories for properly normalized elementary fields in their magnetic dual description. Meanwhile any discrete global symmetries must be present in both the confined and deconfined theories, so will determine the detailed structure of the coupling constants and mixing matrices.  In the example we provide in \S \ref{sec:CH} the interplay between a $Z_3$ permutation symmetry and an effective $Z_2$ symmetry in the confined theory (arising from the confinement scheme) ends up mimicking the behaviour of the more traditional $A_4$ flavour symmetry.  Thus tribimaximal mixing is obtained in the neutrino sector, whereas $Z_3$ is broken by the strong coupling in the quark sector to allow a realistic form for the quark mixing matrix as well.

The more geometrical way to think about strong coupling is, as we have said, via the AdS/CFT correspondence. Indeed, as we will discuss in \S\ref{sec:WED}, there is a simple configuration that can mimick the confinement and discrete symmetries of the model described above. First one can replace each strongly coupled $\SU2$ gauge group with an unknown strongly coupled CFT and consequently with a slice of AdS. The composite quark and lepton states are localized on (or near) the Infra-Red (IR) brane of each throat while the elementary states are localized on (or near) the Ultra-Violet (UV) brane. In order for there to be interactions one of course has to sew the three throats together at the UV end, which leads to a multiple throat background of the type discussed in Ref.\cite{Cacciapaglia:2006tg}, each throat corresponding to a single generation. The $Z_{3}$ symmetry that plays a role in the generation of the tribimaximal mixing of the neutrino sector corresponds to symmetry under cyclical permutations of the throats. In the holographic interpretation this symmetry is spontaneously broken by the strongly coupled sectors. This can be manifest as either different bulk masses in the throats or different throat lengths or both.

\begin{figure}[!t]
\begin{center}
\includegraphics[width=5.5cm]{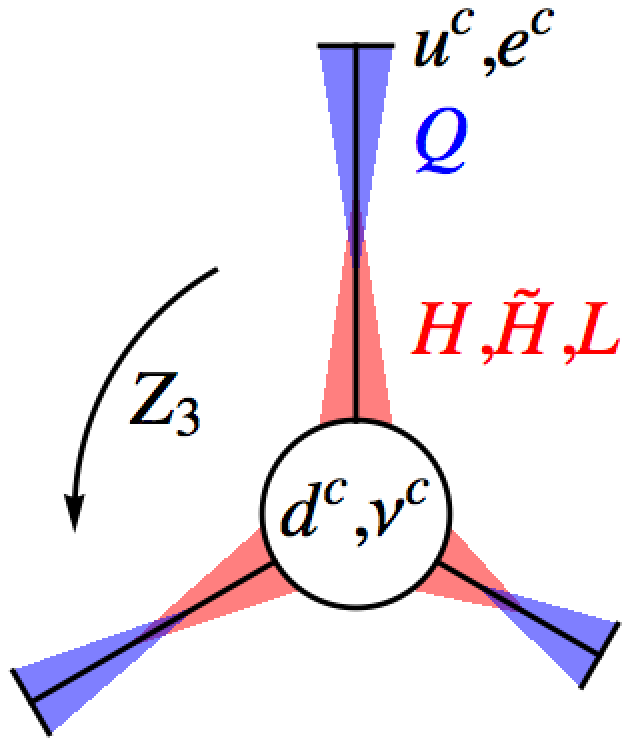}\hspace{1cm}\includegraphics[width=5.5cm]{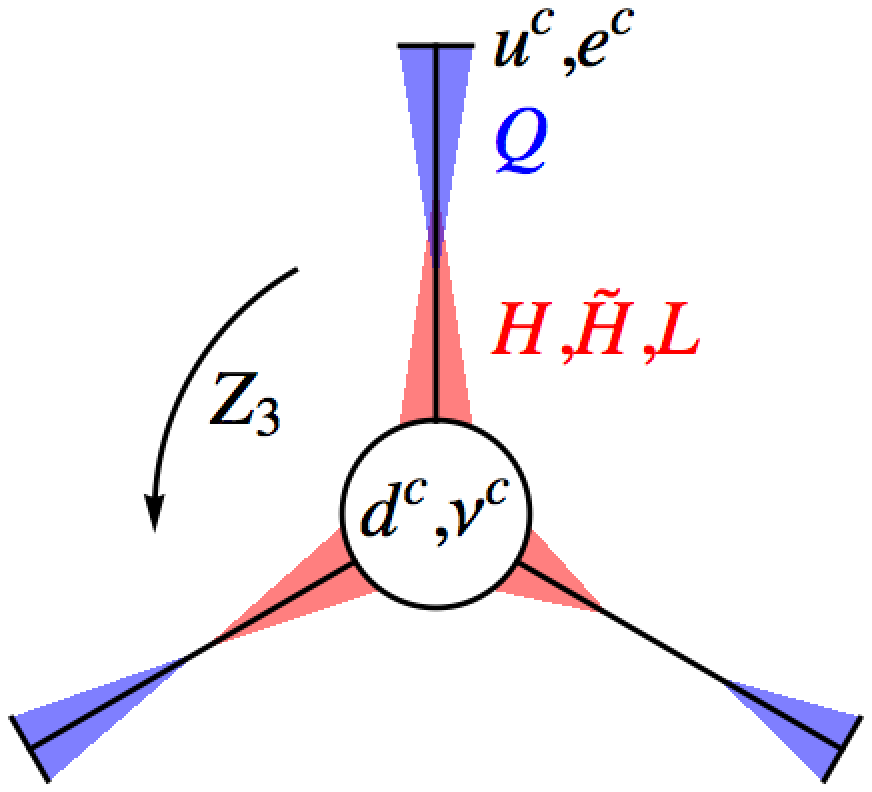}
\caption{The triple confinement picture. Elementary states at the node feel the full discrete symmetry. Bound states are situated at the ends of ``throats". The discrete symmetry is broken either by different throat lengths (left) or by different localisations of the fields in each throat (right), corresponding to different dynamical scales or field normalisations respectively.  Both effects break the discrete symmetry. The picture is also representative of the purely field theoretical configuration.\label{fig:tthroat0}}
\end{center}
\end{figure}

The picture that emerges for the $Z_3$ symmetric case is shown in Figure \ref{fig:tthroat0}. It consistently represents both the field theory and the holographic approaches: it is in fact a completely general geometrical depiction of how to arrange strong coupling hierarchies and a discrete $Z_3$ symmetry in order to achieve the flavour structure observed in nature. In the configuration shown, the neutrinos and quark singlets are mainly elementary and are subject to the discrete symmetry of the UV theory, while the quark doublets and lepton singlets are composite, so their couplings pick up hierarchical factors from the compositeness scales. This is where the $Z_3$ breaking is manifest. The degree of compositeness of a particular state is indicated in Figure \ref{fig:tthroat0} by the distance from the node, which enjoys the full $Z_3$ symmetry. Note that other phenomenologically viable configurations are possible. We should add that a similar arrangement in non-strongly coupled scenarios (for example an orbifold compactification with overlapping wave-functions) would be equally viable (but possibly more difficult to achieve): indeed we regard the present work as essentially putting strong coupling contributions to flavour structure on an equivalent geometrical footing.  An important feature of this diagram is that the up quark Yukawas are composite-composite-elementary couplings, the down quark and charged leptons are composite-elementary-elementary while the neutrino couplings are elementary-elementary-elementary. This explains why the down quarks and charged leptons have similar mass hierarchies, while the ups have roughly the square of the down hierarchy, and the neutrinos have little or no hierarchy and large mixing. We will see that correct mixings are natural and easy to achieve with this configuration.

The geometric picture makes obvious two important additional points. The first is, what is responsible for breaking the $Z_3$ symmetry? How do we for example end up with different throat lengths? In AdS/CFT this can be explained by (for example) invoking the Goldberger-Wise mechanism \cite{Goldberger:1999uk} in which a scalar field acquires a VEV on both the UV and IR branes. In order to generate three different throat lengths, the potential in question must be both $Z_3$ invariant and allow at least three different VEVs for the scalar on the IR brane.  From the 5D point of view the $Z_3$ symmetry is then simply spontaneously broken at tree-level.  From the 4D point of view the VEV represents a coupling (between a source term and composite operators in the CFT) that runs from a UV fixed point to generate alternative forms of radiative symmetry breaking.  The second point that the geometric picture elucidates is what happens to the underlying GUT structure.  In the traditional SU(5) theory $Q,e^c,u^c$ sit in the antisymmetric ${\bf 10}$ and $d^c, L$ sit in the antifundamental ${\bf \overline 5}$.  Therefore the diagram clearly breaks the SU(5) structure; however it is seen to be still {\em approximately} intact, in the sense that the wavefunctions of fields in a given SU(5) multiplet are all localized close to each other in either the UV or the IR. It is reasonable to expect such a slight splitting of wave-functions to occur since, after GUT symmetry breaking, different anomalous dimensions can arise within SU(5) multiplets.

\section{General principles\label{sec:GP}}

Let us first put aside the question of discrete symmetries for a moment, and outline in this section in very broad 
terms how one can take advantage of strong coupling to address flavour structure. We will focus on the ${\cal N}=1$ MSSM throughout\footnote{Although the discussion should in fact also apply to non-supersymmetric models, the dynamical scales we will be using are relatively close to the GUT scale and therefore we assume supersymmetry purely to protect the Higgs mass.}.  For reference we briefly review the objective:  the Yukawa terms determining the flavour structure in the MSSM superpotential are
\be
W_{\rm MSSM}\supset\lambda_{d,ij}d_i^cQ_j\tilde{H}+\lambda_{u,ij}u_i^cQ_jH+\lambda_{e,ij}e_i^cL_j\tilde{H}+\lambda_{\nu,ij}\nu_i^cL_jH\,.
\ee
In order to invoke the see-saw mechanism for neutrino masses there will also be additional mass terms in the neutrino sector which we shall address in due course.  The indices $i$,~$j=1\ldots 3$ label the generations of the matter fields.

Measurements of quark and lepton masses suggest a large hierarchy in both the quark and charged lepton sector, but not necessarily in the neutrino sector.  Specifically, the experimental mass values for the quarks and charged leptons at the weak scale are \cite{Amsler:2008zzb}
\begin{align}\label{eq:exmasses}
m_t & =(169.0-173.3)\mbox{ GeV} & m_b & =(4.13-4.37)\mbox{ GeV} & m_{\tau} & =1.77\mbox{ GeV} \nonumber\\
m_c & =(1.16-1.34)\mbox{ GeV} & m_s & =(70-130)\mbox{ MeV} & m_{\mu} & =106\mbox{ MeV} \\
m_u & =(1.5-3.3)\mbox{ MeV} & m_d & =(3.5-6.0)\mbox{ MeV} & m_e & =0.511\mbox{ MeV}\,. \nonumber
\end{align}
The experimental upper bound on the neutrino masses are 2 eV \cite{Amsler:2008zzb} and the measured mass squared splittings are $\Delta m_{21}^2=7.59\times10^{-5}\mbox{ eV}^2$ and $\Delta m_{32}^2=2.43\times10^{-3}\mbox{ eV}^2$. These two facts suggest neutrino masses in the $(0.1-1)\mbox{ eV}$ range.  Mixing in the weak sector is governed by the CKM matrix: for the present discussion we can represent it in an order of magnitude form, as 
\be\label{eq:CKMex}
V_{\rm CKM}\sim\nbrack{\begin{array}{ccc}
1 & \eta & \varepsilon\eta \\
\eta & 1 & \varepsilon \\
\varepsilon\eta & \varepsilon & 1 \end{array}}
\ee
where $\varepsilon\approx 0.04$ and $\eta\approx 0.22$ (CP violation will for the moment be ignored). The neutrino mixing is described by  the PMNS matrix: the matrix required to diagonalise $\lambda_{\nu,ij}$ with respect to a basis in which $\lambda_{e,ij}$ is already diagonal. Here observations favour the texture of tribimaximal mixing proposed by Harrison, Perkins and Scott \cite{Harrison:2002er}:
\be\label{eq:PMNSex}
V_{\rm PMNS}\approx V_{\rm HPS}=\nbrack{\begin{array}{rrr}
\sqrt{2/3} & \sqrt{1/3} & 0 \\
-\sqrt{1/6} & \sqrt{1/3} & -\sqrt{1/2} \\
-\sqrt{1/6} & \sqrt{1/3} & \sqrt{1/2} \end{array}}.
\ee

To attack these wildly differing flavour structures, we will be appealing to the idea of confinement.  Instead of taking the matter fields of the MSSM to be elementary fields, we assume that some of them are bound states of a strongly coupled sector.  For a given gauge group the dynamical scale corresponds to gauge couplings through RG flow as
\be
\Lambda\sim Ee^{-8b\pi^2/g^2(E)}
\ee
where $b$ is the beta function coefficient for the confining gauge group and $E$ is the renormalisation scale. Therefore they very naturally support an exponentially large hierarchy.  Physically $\Lambda$ is the scale at which the gauge coupling diverges -- i.e.\ the Landau pole scale.  Obviously this idea can explain mass hierarchies for the various MSSM fields, but can it account for the qualitative differences in mixing between the quark and neutrino sectors?  In fact Ref.\cite{Haba:1998wf} has already established that large lepton mixing and small quark mixing can be generated by just a single compositeness scale. We are encouraged by this to seek multiple confinement scale configurations that explain both hierarchies and mixings. Our central proposal is that a mass hierarchy is generated by three separate confining gauge groups, one for each generation: in the discussion of this section the hierarchies will be associated with the confinement scales of the gauge groups\footnote{Compositeness can also generate hierarchies 
if the generations are associated with composite operators of different dimensions as  in  
Ref.\cite{Craig:2009hf} for example.}. 

In order to demonstrate how some of the flavour structure can be generated, suppose that the quark doublets $Q_a$ are bound states of some unknown, strongly coupled gauge group: generically one expects
\be\label{eq:qq}
\Lambda_aQ_a\sim Y_aY_a
\ee
where the $Y_a$ stand for generic (arbitrary and possibly distinct) elementary fields in the deconfined theory, and the dynamical scales $\Lambda_a$ of the strongly coupled theories are the only dimensionful parameters that can make the dimensions of the two operators match.  In ${\cal N}=1$ supersymmetry we can exploit the fact that a strongly coupled theory is well described in the IR by a different, weakly coupled theory through the principle of Seiberg duality \cite{Seiberg:1994pq, Intriligator:1995au} and so identify these bound states and their holomorphic couplings more precisely. The appearance of $\Lambda_a$ with no other small parameters is reasonable in Seiberg duality if we assume that all terms appearing in the K\"ahler potential are of order unity with appropriate powers of the only dimensionful parameter available, $\Lambda_a$.  Let us make the same assumption for the right handed lepton singlets $e^c_a$,
\be\label{eq:ll}
\Lambda_ae^c_a\sim Y_aY_a\, ,
\ee
but \emph{not} for the lepton doublets $L_i$ or right handed neutrino singlets $\nu^c_i$ or up quarks $u^c_i$.  The Yukawa couplings of the low energy theory would appear in the elementary, deconfined theory as higher order nonrenormalizable operators:
\be
W_{\rm UV}\supset\frac{1}{M_X}\nbrack{\xi_{d,ia}d_i^c(Y_aY_a)\tilde{H}+\xi_{u,ia}u_i^c(Y_aY_a)H+\xi_{e,ia}(Y_aY_a)L_i\tilde{H}}+\lambda_{\nu,ij}\nu^c_iL_jH\,,
\ee
where $M_X$ is some high scale and the $\xi$'s and $\lambda_{\nu}$ are assumed to be generic matrices of order unity.  Sums are taken over all indices.  Below the scales $\Lambda_a$ each of the gauge groups confine and the composite operators are replaced with elementary fields using Eqs.~\eqref{eq:qq} and \eqref{eq:ll}, reproducing the Yukawa terms of the MSSM superpotential.  However, a hierarchy now naturally arises in the couplings due to the dynamical scales, i.e.\
\be
\lambda_{ia}\sim\frac{\Lambda_a}{M_X}\xi_{ia}\,.
\ee
Specifically, if we describe the hierarchy in the dynamical scales by two small parameters, $\varepsilon$ and $\eta$, where
\be
\Lambda_{1}=\eta\Lambda_{2}=\varepsilon\eta\Lambda_{3}\, ,
\ee
then the Yukawa couplings for the quarks and charged leptons take the form
\be
\lambda_{ia}\sim\frac{\Lambda_3}{M_X}\nbrack{\begin{array}{ccc}
\varepsilon\eta & \varepsilon & 1 \\
\varepsilon\eta & \varepsilon & 1 \\
\varepsilon\eta & \varepsilon & 1 \end{array}}.
\ee

Note that the masses will not be numerically correct for this simple model since it predicts the up quark and down quark hierarchies to be the same which they clearly are not.  Therefore we cannot expect to reproduce all the features of the Standard Model with this configuration; nevertheless some of the coarser mixing structure is suggestive.  Indeed let us assume (without loss of generality) that our basis for the fields is such that the up quark and charged lepton Yukawas are already diagonal.  In order to diagonalise the down quark Yukawa one makes a bi-unitary rotation:
\be
\hat{\lambda}_d=U_d^{\dag}\lambda_dV_d\,.
\ee
where the matrices $U_d$ and $V_d$ act on the right handed and left handed down quarks respectively.  From $\lambda_d$ it is clear that, while all the elements of $U_d$ will be of order unity, the matrix $V_d$ will take the order of magnitude form
\be\label{eq:VCKMoom}
V_d\sim\nbrack{\begin{array}{ccc}
1 & \eta & \varepsilon\eta \\
\eta & 1 & \varepsilon \\
\varepsilon\eta & \varepsilon & 1 \end{array}}.
\ee
But as $V_d$ (or more precisely $V_d^{\dag}$) rotates the left handed quark doublet it must be the CKM matrix.  Its structure is in keeping with Eq.~\eqref{eq:CKMex}: optimising $m_s$ and taking a running $m_b\sim 2.5$~GeV at the weak scale gives $\eta\sim  0.1$ and $\varepsilon \sim 0.04$. 

Similarly, we can immediately deduce the order of magnitude form of the PMNS matrix.  The neutrino Yukawa is also diagonalised by a bi-unitary rotation
\be
\hat{\lambda}_{\nu}=U_{\nu}^{\dag}\lambda_{\nu}U_{\nu}^{\prime}\,.
\ee
But no composite fields appear in the neutrino Yukawa so the components of $\lambda_{\nu}$ come directly from the UV theory and are expected to be generic and of order unity.  Hence the elements of both $U_{\nu}$'s are also of order unity.  $U_{\nu}$ is identified with the PMNS matrix, with its order one components favouring large neutrino mixing angles.  The general argument would not be drastically changed if we were to allow the right handed quarks and/or neutrinos to be composite in the same way, or if we allowed all three generations of lepton doublets to be composite under a single, strongly coupled gauge group, giving just one scale and thus no hierarchy in the mixing matrices.

This captures the basic structure of both mixing matrices and also gives hierarchical masses for quarks and leptons. However there is clearly some mismatch in the numerical values, so this simple example does not yet have the required detail.  Moreover as already mentioned the experimentally favoured form of the PMNS matrix hints strongly that there is some discrete symmetry at work as well, such as the $A_4$ symmetry of Refs.\cite{A4refs}.  Most examples of discrete symmetries in this context require an extended Higgs sector to achieve the correct mixing, but we will see that compositeness opens up new possibilities.  Indeed by putting the deconfined degrees of freedom (i.e.\ the $Y$'s)  into appropriate representations of some discrete symmetry we can get additional constraints on the $\xi$'s appearing in the UV superpotential of Eq.~\eqref{eq:ll}. If these discrete symmetries are only broken by the strongly coupled sectors of the theory, the constraints remain in the lepton sector of the confined theory.  

In the following section we show how in this way a composite model with a $Z_3$ permutation symmetry can mimick the usual $A_4$ mechanism for the leptons, whilst maintaining the hierarchical quark structure generated by the different confinement scales, along the lines described above.  Many other variations with different discrete symmetry groups are possible but the one we favour has the particularly simple and attractive geometric interpretation described in the Introduction.

\section{Hierarchies driven by confinement scales\label{sec:CH}}

We will now consider an explicit example of the principles discussed in the previous section and embed into it a discrete symmetry.  We will use an SU(5) GUT structure, however this could just be used as a book-keeping device rather than for physical reasons so, although we will include a $\brm{5}$ and $\cbrm{5}$ of Higgs fields, only the doublets need actually be present in the physical theory. The gauge group is extended to include three copies of SU(2), which will make up the confining sector of the model, giving the overall gauge group
\be
\SU{5}\times\SU{2}^3\,.
\ee
In addition, there is a global $Z_3$ permutation symmetry between the three SU(2) gauge groups.  The strong coupling dynamics of the SU(2)'s will ultimately break this symmetry: in this case one could imagine that the $Z_3$ symmetry is spontaneously broken by the VEVs of moduli fields in the underlying theory that determine the SU(2) couplings. An alternative possibility is simply that the SU(2) couplings flow in three different directions from a UV fixed point.

A simple choice for the theory's matter content is given in Table \ref{tab:CHdeconfined}.  The fields $\tilde{P}_i$, $\tilde{Q}_i$  and $Y^a$ are, unlike the Higgses, in complete $\SU{5}$ multiplets in order to preserve gauge coupling unification and ensure anomaly cancellation.  The $Z_3$ permutation symmetry is taken to act on both the generation index $i$ and the SU(2) gauge group label $a$.  There is an $R$-parity that will reduce to the usual matter parity in the low energy theory.  We also use the expedient of a global U(1) symmetry to prevent any operators involving $\tilde{P}$ that could result in proton decay. The choice of charges is not unique and other possibilities can be considered: for example setting the $\tilde{P}_i$ and $Y_a Z_a$ charges to be $+1$ rather than $-1$ and having no $\U1$ would result in the light Higgs being instead a linear combination of these fields and $H$ and $\tilde{H}$.  We discuss this possibility in \S\ref{sec:AH}.

The most general superpotential compatible with the symmetries up to fifth order in the fields\footnote{There are actually some additional operators at fourth order but they remain fourth order in the confined theory so are irrelevant in the IR.} is
\ba\label{eq:CHWuv}
W_{\rm UV} &=& \mu\tilde{H}H+\xi_{P,ia}\tilde{P}_iY_aZ_a+\frac{1}{M_X}\xi_{Q,ia}\tilde{Q}_iY_aY_a\tilde{H}+\frac{1}{M_X^2}\zeta_{ab}\epsilon^{(5)}Y_aY_aY_bY_bH+\nonumber\\
&& \lambda_{\nu,ij}\nu^c_i\tilde{Q}_jH+M_{\nu,ij}\nu^c_i\nu^c_j\,.
\ea
where the $\xi$'s, $\zeta$, $\lambda_{\nu}$ and $M_{\nu}$ are coupling constants and $\epsilon^{(5)}$ denotes a contraction over the SU(5) gauge indices with a rank five alternating tensor.  The coupling constants are constrained by the $Z_3$ symmetry to take the general forms
\be\label{eq:CHxi}
\xi\,,\lambda_{\nu}=\nbrack{\begin{array}{ccc}
a & b & c \\
c & a & b \\
b & c & a \end{array}},\quad
\zeta\,,\frac{1}{m}M_{\nu}=\nbrack{\begin{array}{ccc}
x & y & y \\
y & x & y \\
y & y & x \end{array}}.
\ee
We derive these expressions in Appendix \ref{app:Z3}, where it is shown that $a$, $b$, $c$, $x$ and $y$ are undetermined dimensionless constants (different for each of the matrices) and $m$ is an undetermined mass scale.

Each SU(2) group sees six fundamentals so they all independently $s$-confine at scales $\Lambda_a$.  We will assume these dynamical scales are related hierarchically such that
\be\label{eq:CHepeta}
\varepsilon=\frac{\Lambda_{2}}{\Lambda_{3}}\,,\quad\eta=\frac{\Lambda_{1}}{\Lambda_{2}}
\ee
for some small parameters $\varepsilon$ and $\eta$.  As usual, the $s$-confinement does not change the slope of any of the $\beta$-functions so gauge unification takes place at the usual scale independently of the choice of $\Lambda_a$'s.  We then define elementary IR fields
\be\label{eq:CHmesons}
\Lambda_aA_a\sim Y_aY_a\,,\quad\Lambda_aP_a\sim Y_aZ_a
\ee
to arrive at the matter content of Table \ref{tab:CHconfined} and the superpotential
\ba\label{eq:CHWir}
W_{\rm IR} &=& \mu\tilde{H}H+\Lambda_a\xi_{P,ia}\tilde{P}_iP_a+\frac{\Lambda_a}{M_X}\xi_{Q,ia}\tilde{Q}_iA_a\tilde{H}+\frac{\Lambda_a\Lambda_b}{M_X^2}\xi_{A,ab}\epsilon^{(5)}A_aA_bH+\nonumber\\
&& \lambda_{\nu,ij}\nu^c_i\tilde{Q}_jH+M_{\nu,ij}\nu^c_i\nu^c_j+\epsilon^{(5)}A_aA_aP_a\,.
\ea
Since the confinement scales appear with the ${\bf 10}$'s, much of the hierarchical structure of the quarks and charged leptons will resemble that of the so-called ten-centred models described in Refs. \cite{Strassler:1995ia,Nelson:2000sn}.  The final term in \eqref{eq:CHWir} is generated non-pertubatively\footnote{The non-perturbative nature of this term explains why it does not respect the global U(1) symmetry, which is anomalous under the complete gauge group of the UV theory.} by the SU(2) gauge dynamics but it has little significance in the low energy theory.  Assuming that the components of $\xi_{P,ia}$ in Eq.~\eqref{eq:CHWuv} are generic and of order unity, both $P$ and $\tilde{P}$ gain Dirac masses of order $\Lambda_a$ and are integrated out of the confined low energy theory.  However, as $\tilde{P}$ does \emph{not} appear in any Yukawa terms there is no effective proton decay operator generated by the $P$ sector; indeed the only effect of the $AAP$ term is to provide a decay channel from $P$'s to standard model particles.  With the usual SU(5) matter assignments, the low energy superpotential is thus of the MSSM form:
\be
W_{\rm IR}=\mu\tilde{H}H+\lambda_{d,ia}d_i^cQ_a\tilde{H}+\lambda_{u,ab}u_a^cQ_bH+\lambda_{e,ia}e_a^cL_i\tilde{H}+\lambda_{\nu,ij}\nu^c_iL_jH+M_{\nu,ij}\nu^c_i\nu^c_j\,.
\ee
The coupling constants are read straight from Eqs.~\eqref{eq:CHxi} and \eqref{eq:CHWir} as
\be\label{eq:CHlambda}
\lambda_d\,,\lambda_e=
\frac{\Lambda_3}{M_X}\nbrack{\begin{array}{ccc}
a\varepsilon\eta & b\varepsilon & c \\
c\varepsilon\eta & a\varepsilon & b \\
b\varepsilon\eta & c\varepsilon & a \end{array}},\quad
\lambda_u=\nbrack{\frac{\Lambda_3}{M_X}}^2\nbrack{\begin{array}{ccc}
x\varepsilon^2\eta^2 & y\varepsilon^2\eta & y\varepsilon\eta \\
y\varepsilon^2\eta & x\varepsilon^2 & y\varepsilon \\
y\varepsilon\eta & y\varepsilon & x \end{array}}.
\ee
If we are not using SU(5) as a physical GUT symmetry but merely as a book-keeping device, we are free to assign different values to the $a$'s $b$'s and $c$'s in $\lambda_d$ and $\lambda_e$.

\begin{table}[!tb]
\be
\begin{array}{|l|ccccccc|}\hline
\widerow & H & \tilde{H} & \tilde{Q}_i & \nu^c_i & \tilde{P}_i & Y_a & Z_a \\\hline
\widerow \SU{5} & \fund & \afund & \afund & \brm{1} & \afund & \fund & \brm{1} \\
\widerow \SU{2}_a & \brm{1} & \brm{1} & \brm{1} & \brm{1} & \brm{1} & \fund & \fund \\\hline
\widerow \U{1} & 0 & 0 & 0 & 0 & 1 & 0 & -1 \\
\widerow R_p & 1 & 1 & -1 & -1 & -1 & i & i \\\hline
\end{array}\nonumber
\ee
\caption{\em The deconfined $\SU{5}\times\SU{2}^3$ model with elementary Higgses.  The indices $a$ and $i$ run from 1 to 3 and the bottom two symmetries are global. The $\SU{5}$ gauge group is used for book-keeping purposes only.  In addition, there is a $Z_3$ permutation symmetry acting on indices $i$ and $a$.\label{tab:CHdeconfined}}
\end{table}

\begin{table}[!tb]
\be
\begin{array}{|l|ccccccc|}\hline
\widerow & H & \tilde{H} & \tilde{Q}_i & A_a & \nu^c_i & P_a & \tilde{P}_i \\\hline
\widerow \SU{5} & \fund & \afund & \afund & \asymm & \brm{1} & \fund & \afund \\\hline
\widerow \U{1} & 0 & 0 & 0 & 0 & 0 & -1 & 1 \\
\widerow R_p & 1 & 1 & -1 & -1 & -1 & -1 & -1 \\\hline
\end{array}\nonumber
\ee
\caption{\em The matter content of the confined $\SU{5}\times\SU{2}^3$ model with elementary Higgses.\label{tab:CHconfined}}
\end{table}

The appearance of $\varepsilon$'s and $\eta$'s in the Yukawa matrices does not respect the global $Z_3$ permutation symmetry -- in other words we assume that it is only broken by the confinement scales.  This is a dynamical process driven by the different gauge couplings in each of the confining SU(2) gauge groups.  If the gauge couplings were turned off the SU(2)'s would become global symmetries and $Z_{3}$ would remain.   Since the SU(2)'s are asymptotically free, one sees that this is actually the case in the UV.  

\subsection{Masses and mixings\label{sec:CHmass}}

The most general charged lepton Yukawa coupling is given in Eq.~\eqref{eq:CHlambda}. For simplicity first consider the case where it is diagonal, i.e.\ $b=c=0$ and $a=a_e$.  With this choice, the charged lepton mass matrix is
\be\label{eq:CHml}
m_l=\expect{\tilde{H}}\lambda_e=
\frac{\expect{\tilde{H}}\Lambda_3a_e}{M_X}\nbrack{\begin{array}{ccc}
\varepsilon\eta & 0 & 0 \\
0 & \varepsilon & 0 \\
0 & 0 & 1 \end{array}}.
\ee
We thus predict charged lepton masses
\be\label{eq:CHmlev}
m_{\tau}=a_e\expect{\tilde{H}}\frac{\Lambda_3}{M_X}\,,\quad
m_{\mu}= \varepsilon a_e\expect{\tilde{H}}\frac{\Lambda_3}{M_X}\,,\quad
m_e= \varepsilon\eta a_e\expect{\tilde{H}}\frac{\Lambda_3}{M_X}
\ee
and the ratios of these masses fix the values of $\varepsilon$ and $\eta$ to be
\be\label{eq:CHlep}
\varepsilon=\frac{m_{\mu}}{m_{\tau}}\,,\quad
\eta=\frac{m_e}{m_{\mu}}\,.
\ee
In the neutrino sector, tribimaximal mixing is then ensured by the $Z_3$ permutation symmetry, while a see-saw mechanism produces a light neutrino mass.  The mass matrix for the light neutrinos is attained\footnote{This is actually the result for a diagonal neutrino Yukawa $\lambda_{\nu}$.  Setting $b,c\neq0$ only changes the numerical values of the two light neutrino mass eigenvalues and not the mixing pattern, implying that it is the right handed neutrino Majorana which is the source of the neutrino mixing in this model.} using Eq.~\eqref{eq:CHxi};
\be\label{eq:CHmnu}
m_{\nu}=\expect{H}^2\lambda_{\nu}M_{\nu}^{-1}\lambda_{\nu}^T=
\frac{\expect{H}^2a_{\nu}^2}{m(x-y)(x+2y)}\nbrack{\begin{array}{rrr}
x+y & -y & -y \\
-y & x+y & -y \\
-y & -y & x+y \end{array}}.
\ee
To find the PMNS matrix we diagonalise the neutrino mass matrix;
\be\label{eq:CHmnud}
\hat{m}_{\nu}=V_{\rm HPS}^{\dag}m_{\nu}V_{\rm HPS}=
{\rm diag}\sbrack{\frac{\expect{H}^2a_{\nu}^2}{m(x-y)}\,,\,\frac{\expect{H}^2a_{\nu}^2}{m(x+2y)}\,,\,\frac{\expect{H}^2a_{\nu}^2}{m(x-y)}}
\ee
with $V_{\rm HPS}$ as given by Eq.~\eqref{eq:PMNSex}, i.e.\ the tribimaximal mixing matrix.  On the other hand, if the off diagonal terms in the charged lepton Yukawa of Eq.~\eqref{eq:CHlambda} are allowed to vary from zero by an amount $\sigma$ (which is permitted by the $Z_3$ symmetry) one expects deviations from tribimaximal mixing.  We investigate this possibility more thoroughly in Appendix \ref{app:dtbm} where it is shown that if $\sigma<\varepsilon^2<1$ the deviations are of order $\sigma$ and are otherwise order unity.  The masses of the neutrinos are unaffected. Since $\varepsilon \sim 0.1$, this implies that $\lambda_e$ must be close to diagonal. 
This could be the result of underlying string selection rules, more generally texture zeros, or it may have a direct geometric interpretation as we will see later in the holographic 
incarnation of this model.

Eq.~\eqref{eq:CHmnud} shows that two of the neutrino masses are predicted to be degenerate while the third remains distinct -- in the case where $b=c=0$ one finds
\be\label{eq:CHmnuev}
m_{\nu1}=m_{\nu2}=\frac{\expect{H}^2a_{\nu}^2}{m(x-y)}\,,\quad m_{\nu3}=\frac{\expect{H}^2a_{\nu}^2}{m(x+2y)}\,.
\ee
If $a$, $x$ and $y$ are of order unity and the Higgs VEV is of order the weak scale we must have $m\sim10^{16}\mbox{ GeV}$ to put the scale of the neutrino masses around $\sqrt{\Delta m_{32}^2}\sim0.1\mbox{ eV}$.  The experimental measurement $\Delta m_{32}^2=2.43\times10^{-3}\mbox{ eV}^2$ is then automatically satisfied so long as $x/y\sim1$.  In addition, both normal and inverted hierarchies are readily attained due to the complex nature of $x$ and $y$.  To split the degeneracy between $m_{\nu1}$ and $m_{\nu2}$ and reproduce the observed splitting $\Delta m_{21}^2=7.59\times10^{-5}\mbox{ eV}^2$ one can appeal to radiative corrections similar to those calculated in Ref.\cite{Babu:2002dz}.

Turning to the quark sector, we will first consider a simplified case to illustrate a general, physically significant feature of the model.  However we will soon find that, as in Refs.~\cite{Strassler:1995ia,Nelson:2000sn},
these simplifications give a Cabibbo angle and an electron mass which are both too small, so we will later move to a more general model that gives the correct quark mixing.  Neglecting the CKM matrix for the moment, let us choose $b=c=y=0$ (and $a=a_d$, $x=x_u$) in the quark Yukawa matrices of Eq.~\eqref{eq:CHlambda}. One then has diagonal quark mass matrices with eigenvalues
\begin{align}\label{eq:CHmq}
m_t & =x_u\expect{H}\nbrack{\frac{\Lambda_3}{M_X}}^2 &
m_c & =\varepsilon^2x_u\expect{H}\nbrack{\frac{\Lambda_3}{M_X}}^2 &
m_u & =\varepsilon^2\eta^2x_u\expect{H}\nbrack{\frac{\Lambda_3}{M_X}}^2\nonumber\\
m_b & =a_d\expect{\tilde{H}}\frac{\Lambda_3}{M_X} &
m_s & =\varepsilon a_d\expect{\tilde{H}}\frac{\Lambda_3}{M_X} &
m_d& =\varepsilon\eta a_d\expect{\tilde{H}}\frac{\Lambda_3}{M_X}\,.
\end{align}
The model therefore predicts a relationship between mass ratios in the up and down quark sectors:
\be\label{eq:CHqep}
\frac{m_c}{m_t}=\nbrack{\frac{m_s}{m_b}}^2=\varepsilon^2\,,\quad
\frac{m_u}{m_c}=\nbrack{\frac{m_d}{m_s}}^2=\eta^2\,.
\ee
Comparing with the experimental masses in Eq.~\eqref{eq:exmasses} we find that both ratios are in agreement with the data (up to factors of order unity) provided one uses the expected running bottom mass of around $2.5\mbox{ GeV}$.  This suggests an explanation for the difference in mass hierarchies between the up quark and down quark sector: the down quark Yukawa comes from a composite-elementary-elementary operator whereas the up quark Yukawa comes from a composite-composite-elementary operator, meaning the lower generations of up quarks acquire an extra suppression in their mass terms and therefore that the mass hierarchy is more extreme (i.e.\ roughly the square of the down quark and charged lepton hierarchies).

The expressions \eqref{eq:CHqep} for $\varepsilon$ and $\eta$ required in the quark sector appear to be at odds with the ratios \eqref{eq:CHlep} required for the lepton sector.  This is in fact the second feature that the model shares with the ten-centred models.
One can account for this difference by invoking the unknown field normalisations.  The fields $e^c_a$, $u^c_a$ and $Q_a$ all come from composite operators defined in Eq.~\eqref{eq:CHmesons}, but this expression does \emph{not} specify the normalisation of the fields arising from the K\"ahler potential. As we already mentioned, in Seiberg duality it is reasonable to take the factors to be of order unity, but strictly speaking one has $\Lambda_aA_a=\alpha_{A,a}Y_aY_a$ for some undetermined and incalculable constant $\alpha$.  Since the SU(5) gauge group is ultimately broken, it is conceivable that different normalisation constants 
arise for $e^c_a$, $u^c_a$ and $Q_a$.  The net effect is to allow different effective values of $\varepsilon$ and $\eta$ in each Yukawa coupling, i.e.\
\begin{align}\label{eq:CHvepeta}
\varepsilon_{u}= & \sqrt{\frac{\alpha_{Q2}\alpha_{u2}}{\alpha_{Q3}\alpha_{u3}}}\varepsilon & \varepsilon_{d}= & \frac{\alpha_{Q2}}{\alpha_{Q3}}\varepsilon & \varepsilon_{e}= & \frac{\alpha_{e2}}{\alpha_{e3}}\varepsilon & \nonumber\\
\eta_{u}= & \sqrt{\frac{\alpha_{Q1}\alpha_{u1}}{\alpha_{Q2}\alpha_{u2}}}\eta & \eta_{d}= & \frac{\alpha_{Q1}}{\alpha_{Q2}}\eta & \eta_{e}= & \frac{\alpha_{e1}}{\alpha_{e2}}\eta & 
\end{align}
with $\varepsilon$ and $\eta$ defined as in Eq.~\eqref{eq:CHepeta}.  If we desire the field normalisations to be natural and of order unity (allowing values of $\alpha$ around $1/3-3$ say) such factors are sufficient to allow consistency between Eqs.~\eqref{eq:CHlep} and \eqref{eq:CHqep}.

Now we turn to the CKM matrix and allow the down quark mass matrix to be off-diagonal, $b,~c\neq0$ (allowing both mass matrices to be off diagonal does not change things significantly). Alas it can be shown that the CKM matrix cannot be reproduced satisfactorily.  For the $Z_3$ imposed structure of Eq.~\eqref{eq:CHxi}, it is always diagonal in the degenerate limit\footnote{This can be seen by rotating the both the left and right handed quark multiplets into their $Z_3$ eigenstates upon which the matrices $\xi_Q$ and $\zeta$ are simultaneously diagonalised.} $\varepsilon=\eta=1$ -- any off diagonal corrections are generated by the quark mass ratios.  The problem therefore lies with an incompatibility between the mass ratio relations of Eq.~\eqref{eq:CHqep} and the size of the Cabibbo angle: the Cabibbo angle is always too small when the mass ratio relations are satisfied.  One can see why by referring to \S\ref{sec:GP}.  The arguments there suggest that, if the mass ratio relations are satisfied
\be
V_{\rm CKM}\sim\nbrack{\begin{array}{ccc}
1 & \eta & \varepsilon\eta \\
\eta & 1 & \varepsilon \\
\varepsilon\eta & \varepsilon & 1 \end{array}}\sim\nbrack{\begin{array}{ccc}
1 & \sqrt{m_u/m_c} & \sqrt{m_u/m_t} \\
\sqrt{m_u/m_c} & 1 & \sqrt{m_c/m_t} \\
\sqrt{m_u/m_t} & \sqrt{m_c/m_t} & 1 \end{array}}.
\ee
Though most of the entries can be made acceptable by varying the parameters in $\xi_Q$ and $\zeta$, this form always leads to values of $|V_{us}|$ and $|V_{dc}|$ (i.e.\ the Cabibbo angle) that are too small by a factor of five to ten.

However, we can again address this problem by considering the field normalisations.  Indeed, it is easily shown that the Yukawas of Eq.~\eqref{eq:CHlambda} with hierarchies defined by Eq.~\eqref{eq:CHvepeta} can reproduce the flavour structure of the MSSM entirely, including any CP violation.  For example, choosing a diagonal up quark Yukawa and a down quark Yukawa parameterised by $\varepsilon_d=0.051$, $\eta_d=0.19$ and $(a,b,c)=(e^{i\pi/3},3e^{5i\pi/3},3e^{4i\pi/3})$ results in
\be
\frac{m_s}{m_b}=0.043\,,\quad\frac{m_d}{m_s}=0.084\,,\quad
|V_{\rm CKM}|=\nbrack{\begin{array}{ccc}
0.986 & 0.168 & 0.005 \\
0.168 & 0.985 & 0.023 \\
0.009 & 0.027 & 1.000 \end{array}},\quad
J=2.14\times10^{-5}\,.
\ee
Here, $J$ is the Jarlskog invariant \cite{Jarlskog:1985cw, Amsler:2008zzb} which has experimental value $\nbrack{3.05^{+0.19}_{-0.20}}\times10^{-5}$ and is non-zero if and only if CP violation is present in the quark sector (see Appendix \ref{app:J}).  To satisfy the experimental constraints on the up quark and charged lepton masses one then needs to choose field normalisations
\be
\frac{\alpha_{e2}}{\alpha_{e3}}\approx0.4\frac{\alpha_{u2}}{\alpha_{u3}}\approx\frac{\alpha_{Q2}}{\alpha_{Q3}}
\sep{and}
\frac{\alpha_{e1}}{\alpha_{e2}}\approx0.5\frac{\alpha_{u1}}{\alpha_{u2}}\approx0.03\frac{\alpha_{Q1}}{\alpha_{Q2}}\,.
\ee
Therefore an additional small hierarchy in the normalisation of the first two generations is desirable, although strictly speaking we can get away with only order unity parameters by setting $\alpha_{Q1}\approx3$ and $\alpha_{Q2}\approx1/3$.

The final aspects of the model we will examine are the absolute mass scales of the various fields.  Combining the down quark Yukawa \eqref{eq:CHlambda} with the up quark and charged lepton masses of Eqs.~\eqref{eq:CHmq} and \eqref{eq:CHmlev} one finds expressions for the third generation masses
\be
m_t\sim\expect{H}\nbrack{\frac{\Lambda_3}{M_X}}^2\,,\quad
m_b\sim m_{\tau}\sim\expect{\tilde{H}}\frac{\Lambda_3}{M_X}
\ee
leading to the relations
\be
\Lambda_3\sim\frac{m_t}{m_{\tau}\tan\beta}M_X
\ee
between the fundamental scales of the model, where $\tan\beta=\expect{H}/\expect{\tilde{H}}$.  One can see that this model favours a large value of $\tan\beta$ (greater than about 10 for reasonable values of $a_e$ and $x_u$) and similar values for the two scales; $\Lambda_3\sim M_X$.  In fact, it is economical to choose the right handed neutrino mass scale to be around $M_X$ too.  All new fundamental high scales would then be around the GUT scale of $10^{16}\mbox{ GeV}$, with only the usual MSSM parameter $\mu\sim1\mbox{ TeV}$ being different.  Other than this argument of economy there is no immediate constraint on the absolute values of $M_{X}$ and $\Lambda_{3}$, only their ratio.  One could therefore imagine a UV completion similar to Ref.~\cite{Nelson:1996km} where there is no need to to take $M_{X}$ to be as high as the GUT scale.  In this scenario the only constraints on the compositeness scale would be the non-observation of the additional fields $P$ and $\tilde{P}$ (charged under the SM gauge group) and the apparently elementary nature of the SM fermions up to the TeV scale -- it is entirely feasible that the compositeness scale could be as low as a few TeV.

\section{Hierarchies driven by anomalous dimensions\label{sec:AH}}

We have just seen how hierarchies in the MSSM can be introduced by the dynamical scales of multiple strongly coupled gauge groups.  An alternative possibility is to use the the anomalous dimensions of various operators and the resulting wavefunction renormalisation.  The general idea was introduced in Ref.~\cite{Nelson:2000sn} and can be summarised as follows.  Suppose we take the coupling
\be
W_{\rm IR}\supset\frac{1}{M_{X}}\xi_{ia}\tilde{Q}_{i}Y_{a}Y_{a}\tilde{H}
\ee
and investigate the RG flow of the coupling constant $\xi$.  Wavefunction renormalisation is the only effect we need to consider in supersymmetric field theories, so we find
\be
\beta_{\xi}= \xi+\frac{1}{2}\xi(\gamma_{\tilde{Q}}+2\gamma_{Y}+\gamma_{\tilde{H}})
\ee
where the $\gamma$'s are the anomalous dimensions of the elementary fields appearing in the Yukawa couplings, i.e.\
\be
{\rm dim}Y=1+\frac{1}{2}\gamma_{Y}\,,\quad\mbox{etc.}
\ee
Ergo if the Yukawa coupling is defined at some high scale $M_{X}$ -- where its value is naturally of order unity -- and the anomalous dimensions remain roughly constant, one finds that it runs to
\be
\xi\sim\nbrack{\frac{\Lambda}{M_{X}}}^{1+\frac{1}{2}(\gamma_{\tilde{Q}}+2\gamma_{Y}+\gamma_{\tilde{H}})}
\ee
at some lower scale $\Lambda$.  Now if we allow the anomalous dimensions to be \emph{different} for each generation we can easily arrange for a large hierarchy in the Yukawa couplings, as long as $\Lambda/M_{X}$ is sufficiently small and the anomalous dimensions are sufficiently large.  In particular we could assume that $\gamma_{\tilde{Q}1}=\gamma_{\tilde{Q}2}=\gamma_{\tilde{Q}3}$ to find
\be
\xi_{ia}\sim\xi_{0}\nbrack{\frac{\Lambda}{M_{X}}}^{\gamma_{Y,a}}
\ee
which generically reproduces the hierarchical structure\footnote{An interesting variation on this idea would be to have negative anomalous dimensions, with those for the third generation being furthest from zero.  The Yukawa couplings of the third generation would then be \emph{enhanced} by RG flow, rather than the Yukawa couplings of the first generation being suppressed as is usually the case.  Such an approach actually seems to fit better with the observed mass hierarchies: the masses of the first generation particles are very similar whereas there are increasingly large discrepancies in the masses of subsequent generations.} found in \S\ref{sec:CH} when $\gamma_{Y1}>\gamma_{Y2}>\gamma_{Y3}$.

With this in mind we will now show how a simpler model\footnote{The model is based on Ref.~\cite{Abel:2009bj}, where a confined $\SU{5}$ GUT model was introduced as the magnetic description of an electric dual GUT -- an $\SU{11}\times\SU{2}^3$ theory for example.  In this paper we shall instead assume the deconfined magnetic theory to be valid upto the GUT scale.} can be constructed, where the hierarchies are driven only by the anomalous dimensions of the elementary fields in the strongly coupled sector.  It should, however, be noted that the arguments below would apply equally validly to the model presented in \S\ref{sec:CH}.  The deconfined theory will be an $\SU{5}\times \SU{2}^3$ model with a $Z_{3}$ permutation symmetry, as in \S\ref{sec:CH}, with each $\SU{2}$ corresponding to a single generation.  Rather than introducing any additional composite states we will associate the $Y_{a}Z_{a}$ operators with the MSSM Higgs fields.  Of course, this means there will now be three generations of Higgses $H_{a}$, and also that $\tilde{H}$ should be appended with a flavour index so that all Higgses get masses and the gauge anomalies cancel.  The complete matter content is shown in Tables \ref{tab:AHdeconfined} and \ref{tab:AHconfined}; the deconfined and confined theories respectively.  Even if the SU(5) gauge group is being used for notational convenience rather than any physical reason, the Higgs triplets are unavoidable due to the nature of the $H_{a}$ as composite operators.  The $R$-parity acts as before but there is no longer any need for an extra U(1) symmetry.  Note that as far as the confining SU(2)'s are concerned the matter content is identical to \S\ref{sec:CH} so $s$-confinement is unchanged.

\begin{table}[!tb]
\be
\begin{array}{|l|ccccc|}\hline
\widerow & \tilde{H}_{i} & \tilde{Q}_i & \nu^c_i & Y_a & Z_a \\\hline
\widerow \SU{5} & \afund & \afund & \brm{1} & \fund & \brm{1} \\
\widerow \SU{2}_a & \brm{1} & \brm{1} & \brm{1} & \fund & \fund \\\hline
\widerow R_p & 1 & -1 & -1 & -i & i \\\hline
\end{array}\nonumber
\ee
\caption{\em The deconfined $\SU{5}\times\SU{2}^3$ model with a composite Higgs.  The indices $a$ and $i$ run from 1 to 3 and the bottom two symmetries are global. The $\SU{5}$ gauge group is used for book-keeping purposes only.  In addition, there is a $Z_3$ permutation symmetry acting on indices $i$ and $a$.\label{tab:AHdeconfined}}
\end{table}

\begin{table}[!tb]
\be
\begin{array}{|l|ccccc|}\hline
\widerow & H_{a} & \tilde{H}_{i} & \tilde{Q}_i & A_a & \nu^c_i \\\hline
\widerow \SU{5} & \fund & \afund & \afund & \asymm & \brm{1} \\\hline
\widerow R_p & 1 & 1 & -1 & -1 & -1 \\\hline
\end{array}\nonumber
\ee
\caption{\em The matter content of the confined $\SU{5}\times\SU{2}^3$ model with a composite Higgs.\label{tab:AHconfined}}
\end{table}

The superpotential we will use in the deconfined theory is similar to Eq.~\eqref{eq:CHWuv} and reads
\be
W_{\rm UV}=\xi_{H,ia}\tilde{H}_iY_aZ_a+\frac{1}{M_X}\xi_{Q,ija}\tilde{Q}_iY_aY_a\tilde{H}_{j}+\frac{1}{M_{X}}\xi_{\nu,ija}\nu^c_i\tilde{Q}_jY_{a}Z_{a}+M_{\nu,ij}\nu^c_i\nu^c_j\,.
\ee
This is still the most general superpotential consistent with the symmetries of the theory, but now only up to fourth order in the deconfined degrees of freedom.  In order to reproduce the correct flavour structure in this model, we will need to include a second $Z_{3}$ symmetry in the UV theory that acts in the same way but only on $Z$ and $\tilde{H}$.  Thus the matter fields couple only to the singlet states
\be
\tilde{H}_{S}\equiv\tilde{H}_{1}+\tilde{H}_{2}+\tilde{H}_{3}\sep{and}(YZ)_{S}\equiv Y_{1}Z_{1}+Y_{2}Z_{2}+Y_{3}Z_{3}
\ee
and the superpotential becomes
\be
W_{\rm UV}=\xi_{H,ia}\tilde{H}_iY_aZ_a+\frac{1}{M_X}\xi_{Q,ia}\tilde{Q}_iY_aY_a\tilde{H}_{S}+\frac{1}{M_{X}}\xi_{\nu,ij}\nu^c_i\tilde{Q}_j(YZ)_{S}+M_{\nu,ij}\nu^c_i\nu^c_j\,.
\ee
Equivalent symmetry arguments to those found in \S\ref{sec:CH} then give the form \eqref{eq:CHxi} for the elementary couplings $\xi$ and $M_{\nu}$.  We now flow to the IR where the SU(2) factors become strongly coupled and undergo $s$-confinement.  During this process one expects the various couplings in the superpotential to run due to the effect of wavefunction renormalisation, with the running governed by the anomalous dimensions of the elementary fields.  We generally expect the fields to have small anomalous dimensions, but this is not necessarily the case in the strongly coupled sector.  Here the fields are subject to highly non-trivial, non-perturbative effects that can push the anomalous dimensions far from zero.  In addition, the anomalous dimensions depend on non-holomorphic interactions so it is reasonable to assume that they are less constrained by the flavour symmetries appearing in the superpotential.  Consequently we assume that this is where the $Z_{3}$ permutation symmetry is likely to be broken.  Equivalently, we can say that the theory starts at a UV fixed point with a $Z_{3}$ flavour symmetry.  It is then the flow from this fixed point that breaks $Z_{3}$.  We will come back to this idea in \S\ref{sec:WED}.

Supposing this is the case we find
\be
{\rm dim}Y_{a}Y_{a}=2+\frac{1}{2}\gamma_{A,a}\,,\quad
{\rm dim}Y_{a}Z_{a}=2+\frac{1}{2}\gamma_{H,a}
\ee
for some positive, order unity anomalous dimensions $\gamma$ with all other anomalous dimensions vanishing.  This implies that the couplings run as
\ba
\xi_{H,ia} &\longrightarrow& \xi_{H,ia}\nbrack{\frac{\Lambda_{a}}{M_{X}}}^{\frac{1}{2}\gamma_{H,a}}\nonumber\\
\xi_{Q,ija} &\longrightarrow& \xi_{Q,ija}\nbrack{\frac{\Lambda_{a}}{M_{X}}}^{1+\frac{1}{2}\gamma_{A,a}}\nonumber\\
\xi_{\nu,ija} &\longrightarrow& \xi_{\nu,ija}\nbrack{\frac{\Lambda_{a}}{M_{X}}}^{1+\frac{1}{2}\gamma_{H,a}}
\ea
Defining the elementary IR fields
\be\label{eq:AHmesons}
\Lambda_aA_a\sim Y_aY_a\,,\quad\Lambda_aH_a\sim Y_aZ_a
\ee
we thus find an IR superpotential
\ba\label{eq:AHWir}
W_{\rm IR} &=& \Lambda_{a}\nbrack{\frac{\Lambda_{a}}{M_{X}}}^{\frac{1}{2}\gamma_{H,a}}\xi_{H,ia}\tilde{H}_iH_a+\frac{\Lambda_a}{M_X} \nbrack{\frac{\Lambda_{a}}{M_{X}}}^{\frac{1}{2}\gamma_{A,a}}\xi_{Q,ia}\tilde{Q}_iA_a\tilde{H}_{S}+\nonumber\\
&& \frac{\Lambda_a}{M_X} \nbrack{\frac{\Lambda_{a}}{M_{X}}}^{\frac{1}{2}\gamma_{H,a}}\xi_{\nu,ij}\nu^c_i\tilde{Q}_jH_{S}+M_{\nu,ij}\nu^c_i\nu^c_j+A_aA_aH_{a}\,.
\ea
This time the up quark Yukawa -- the $AAH$ term -- is generated non-peturbatively so warrants a more detailed discussion.  The idea that the up quark Yukawa couplings can be generated by $s$-confinement of Sp($M$) gauge groups was noted in Refs.~\cite{Strassler:1995ia, Kitazawa:2004nf, Abel:2009bj}.  The general form of the dynamical superpotential is given by the Pfaffian of the mesons to which the confining group couples:
\be W_{\rm dyn}=-\frac{{\rm Pf}_F({\rm mesons})}{\Lambda_a^{F-3}}
\ee
where $F=2M+4=6$ denotes the effective number of flavours seen by each $s$-confining Sp(1) gauge group.  In the present case these flavours correspond to $Y_a$'s in fundamentals of $\SU{5}$ and singlets $Z_a$. Hence the coupling for each gauge group is \cite{Abel:2009bj}
\be
W_{\rm dyn}=-\frac{\varepsilon^{(5)}(Y_{a}Z_{a})(Y_{a}Y_{a})(Y_{a}Y_{a})}{\Lambda_a^{F-3}}\,.
\ee
Identifying the elementary IR fields as in Eq.~\eqref{eq:AHmesons} we thus find the stated coupling.  It is important to note that the dynamical superpotential does not respect any anomalous global symmetries.  Of particular relevance here is the additional $Z_{3}$ symmetry, which we added to restrict the matter fields to couplings involving only the $Z_{3}$ singlet Higgs states $H_{S}$ and $\tilde{H}_{S}$.  Hence the up quarks are free to couple to \emph{all} components of the Higgs field.  This will be important when calculating the form of the up quark Yukawa coupling.

Adopting the opposing strategy from \S\ref{sec:CH} we will now assume all three dynamical scales are equal and only the anomalous dimensions are distinct.  To simplify the discussion we will reparameterise the hierarchy as before, assuming $\gamma_{1}>\gamma_{2}>\gamma_{3}$ and setting
\begin{align}\label{eq:AHepeta}
\nbrack{\frac{\Lambda}{M_{X}}}^{\frac{1}{2}(\gamma_{H2}-\gamma_{H3})} & =\varepsilon_{H} &
\nbrack{\frac{\Lambda}{M_{X}}}^{\frac{1}{2}(\gamma_{H1}-\gamma_{H2})} & =\eta_{H}\nonumber\\
\nbrack{\frac{\Lambda}{M_{X}}}^{\frac{1}{2}(\gamma_{A2}-\gamma_{A3})} & =\varepsilon_{A} &
\nbrack{\frac{\Lambda}{M_{X}}}^{\frac{1}{2}(\gamma_{A1}-\gamma_{A2})} & =\eta_{A}\,.
\end{align}
The Higgs mass matrix is therefore of the form
\be
\mu_{ia}=\Lambda\nbrack{\frac{\Lambda}{M_{X}}}^{\frac{1}{2}\gamma_{H,a}}\xi_{H,ia}=
h\Lambda\nbrack{\begin{array}{ccc}
a\varepsilon_{H}\eta_{H} & b\varepsilon_{H} & c \\
c\varepsilon_{H}\eta_{H} & a\varepsilon_{H} & b \\
b\varepsilon_{H}\eta_{H} & c\varepsilon_{H} & a \end{array}}
\ee
where $h=(\Lambda/M_{X})^{\gamma_{H3}/2}$.  Generically the eigenvalues are of order $(h\Lambda,\,\varepsilon_{H}h\Lambda,\,\varepsilon_{H}\eta_{H}h\Lambda)$ but if we make the special choice
\be
a+b+c=0
\ee
the smallest eigenvalue becomes zero.  Thus with the usual fine tuning (specifically, one requires $\delta a/a=\mu_{\rm MSSM}/\varepsilon_{H}\eta_{H}h\Lambda$) we can arrange to have one light eigenvalue of around $\mu_{\rm MSSM}$.  Recalling that the SU(5) structure is simply for convenience, we are free to have this fine tuning only in the Higgs doublet mass matrix leaving the triplet masses all above $\varepsilon_{H}h\Lambda$.  As long as $\varepsilon_{H}h\Lambda>M_{\rm GUT}$ the extra Higgs fields do not spoil gauge unification and if $\varepsilon_{H}h\Lambda$ is large enough proton decay will also be suppressed.

At low energy one then goes to the diagonal basis for the Higgses and integrates out all of the triplets and the two heavy generations of doublet.  The light Higgs doublet, $H$, is made up of a linear combination
\be
H\sim H_{1}+\varepsilon_{H}H_{2}+\varepsilon_{H}\eta_{H}H_{3}
\ee
whereas $\tilde{H}$ is made up of a roughly equal combination of $\tilde{H}_{1}$, $\tilde{H}_{2}$ and $\tilde{H}_{3}$.  This means
\ba\label{eq:AHHdef}
(H_{1},\,H_{2},\,H_{3}) & \supset & (H,\,\varepsilon_{H}H,\,\varepsilon_{H}\eta_{H}H) \nonumber\\
(\tilde{H}_{1},\,\tilde{H}_{2},\,\tilde{H}_{3}) & \supset & (\tilde{H},\,\tilde{H},\,\tilde{H})
\ea
which further implies that
\be
H_{S}\equiv H_{1}+H_{2}+H_{3}\supset H\sep{and}\tilde{H}_{S}\equiv\tilde{H}_{1}+\tilde{H}_{2}+\tilde{H}_{3}\supset\tilde{H}
\ee
so we once more find a low energy superpotential of the MSSM form
\be
W_{\rm IR}=\mu\tilde{H}H+\lambda_{d,ia}d_i^cQ_a\tilde{H}+\lambda_{u,ab}u_a^cQ_bH+\lambda_{e,ia}e_a^cL_i\tilde{H}+\lambda_{\nu,ij}\nu^c_iL_jH+M_{\nu,ij}\nu^c_i\nu^c_j\,.
\ee
Since the $Z_{3}$ singlet Higgs states $H_{S}$ and $\tilde{H}_{S}$ both contain order unity components of the light Higgs doublets $H$ and $\tilde{H}$ we can simply replace $H_{S}\rightarrow H$ and $\tilde{H}_{S}\rightarrow\tilde{H}$ in the down quark and lepton Yukawas.  Applying the parameterisation defined in Eq.~\eqref{eq:AHepeta}, most Yukawa couplings can then be read straight from Eq.~\eqref{eq:AHWir} to be
\be
\lambda_{d}\sim\lambda_{e}\sim\frac{\Lambda}{M_X}\nbrack{\begin{array}{ccc}
a\varepsilon_{A}\eta_{A} & b\varepsilon_{A} & c \\
c\varepsilon_{A}\eta_{A} & a\varepsilon_{A} & b \\
b\varepsilon_{A}\eta_{A} & c\varepsilon_{A} & a \end{array}},\quad
\lambda_{\nu,ij}\sim\xi_{\nu,ia}\,.
\ee
We therefore find \emph{exactly} the same form for the down quark and lepton Yukawa couplings as given in Eqs.~\eqref{eq:CHxi} and \eqref{eq:CHlambda}.  On the other hand, mixing between the $Z_{3}$ Higgs eigenstates in the up quark sector produces a different up quark Yukawa.  Including all three components of the Higgs field, the up quark Yukawa couplings take the diagonal form
\be
u^{c}_{1}Q_{1}H_{1}+u^{c}_{2}Q_{2}H_{2}+u^{c}_3Q_{3}H_{3}
\ee
where each term can be multiplied by an order unity coupling constant.  Upon integrating out the heavy Higgs fields and using the relations of Eq.~\eqref{eq:AHHdef} we thus find
\be
\lambda_{u}\sim\nbrack{\begin{array}{ccc}
1 & 0 & 0 \\
0 & \varepsilon_{H} & 0 \\
0 & 0 & \varepsilon_{H}\eta_{H} \end{array}}.
\ee

If we choose $\varepsilon_{H}=\varepsilon_{A}^{2}$ and $\eta_{H}=\eta_{A}^{2}$ the only real difference between this model and the model in \S\ref{sec:CH} is that the up quark Yukawa is forced to be diagonal.  In other words, all of the arguments about masses mixing given in \S\ref{sec:CHmass} are directly applicable to this model.  As with the previous model we are free to vary the $\alpha$'s independently for the up quark singlet, quark doublet and charged lepton singlet to reach the correct mass and mixing parameters, although this time we have more freedom as we can vary the relationship between $\varepsilon_{H},\,\eta_{H}$ and $\varepsilon_{A},\,\eta_{A}$ as well.  The exception to the comparison is with the absolute masses of the various particles.  Since the up quark Yukawas come from a non-peturbative term they are not suppressed by the high scale $M_{X}$, whereas the down quark Yukawas and charged lepton Yukawas are.  Ergo it is more natural for the top quark mass to be large in this model, which was not the case previously.  Quantitatively we require
\be
\frac{\Lambda}{M_{X}}\sim\frac{m_{b}}{m_{t}}\tan\beta
\ee
(with $\Lambda>M_{\rm GUT}$ to suppress the Higgs triplet masses).  A sensible choice would therefore be $M_{X}=M_{\rm Pl}$ which leaves a large range for $\Lambda$ in which proton decay from our heavy Higgs triplets is not problematic.

Despite their mathematical similarities, the physical interpretation of this model is quite different.  Although the $Z_{3}$ symmetry breaking and mass hierarchy patterns are the same as in \S\ref{sec:CH}, they are now being driven by the anomalous dimensions of the composite operators rather than the dynamical scales of the theory.  Nonetheless both effects are a direct result of strongly coupled physics.  

\section{Geometric realisation\label{sec:WED}}

The models presented in the previous section were in summary ten-centred models in which the hierarchies were indeed generated by compositeness, and in which {\em either} the confinement scales {\em or} the anomalous dimensions broke an underlying discrete symmetry. The neutrinos then indeed retained the discrete symmetry of the UV theory because their couplings did not involve any composite fields.  It is natural try to ask if one can have configurations in which the discrete symmetry is broken partly by the confinement scales and partly the anomalous dimensions. In particular this may help to alleviate some of the problems we found with the small Cabibbo angle. 
In order to realise these more general configurations a geometric picture is far more flexible: it will also neatly encapsulate the relation between the discrete symmetry  and the strong coupling. 

In order to construct a geometric realisation, it is natural to use the AdS/CFT correspondence.  In particular, models in which there are a number of strongly coupled conformal sectors coupled to a weakly interacting set of elementary fields is naturally described by a multiple-throat geometry of the kind analysed in Ref.\cite{Cacciapaglia:2006tg}.  For a three throat model we can incorporate a $Z_3$ permutation symmetry as a rotational symmetry amongst the throats.  The $Z_3$ breaking by different dynamical scales then corresponds to different throat lengths.  From the 5D point of view this is simply the spontaneous breaking of the discrete symmetry by the Goldberger-Wise mechanism \cite{Goldberger:1999uk} (i.e.\ one would write down stabilizing superpotentials on the UV and IR branes for a bulk scalar field that can give at least three different possible minima on the IR branes, leading to spontaneous breaking of the $Z_3$ permutation symmetry).  The holographic interpretation is that the conformal symmetry of the 4D theory is broken spontaneously, and that this breaking can happen at different dynamical scales.  The alternative model incorporated $Z_3$ breaking by different anomalous dimensions, and this corresponds holographically to different bulk masses in exactly the same setup.

\begin{figure}[!t]
\begin{center}
\includegraphics[width=4cm]{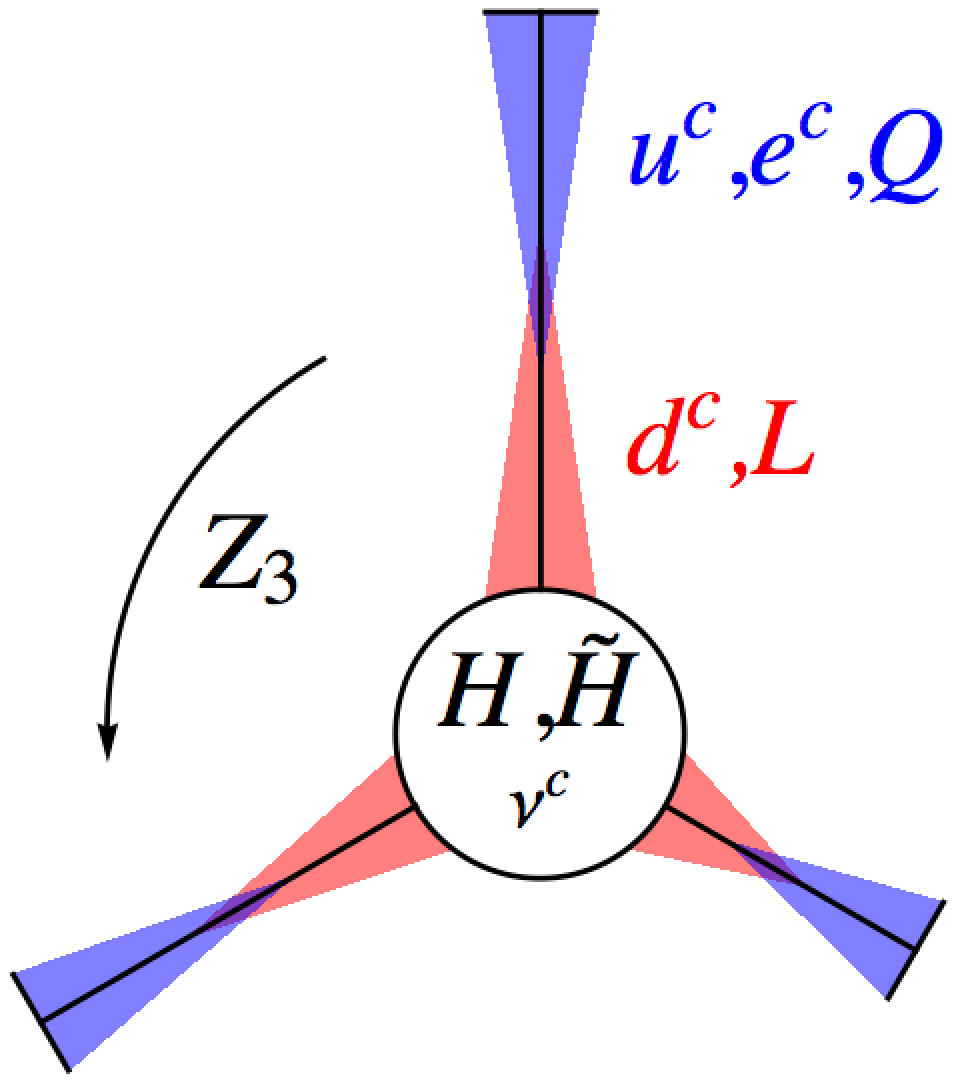}\hspace{2cm}
\includegraphics[width=6cm]{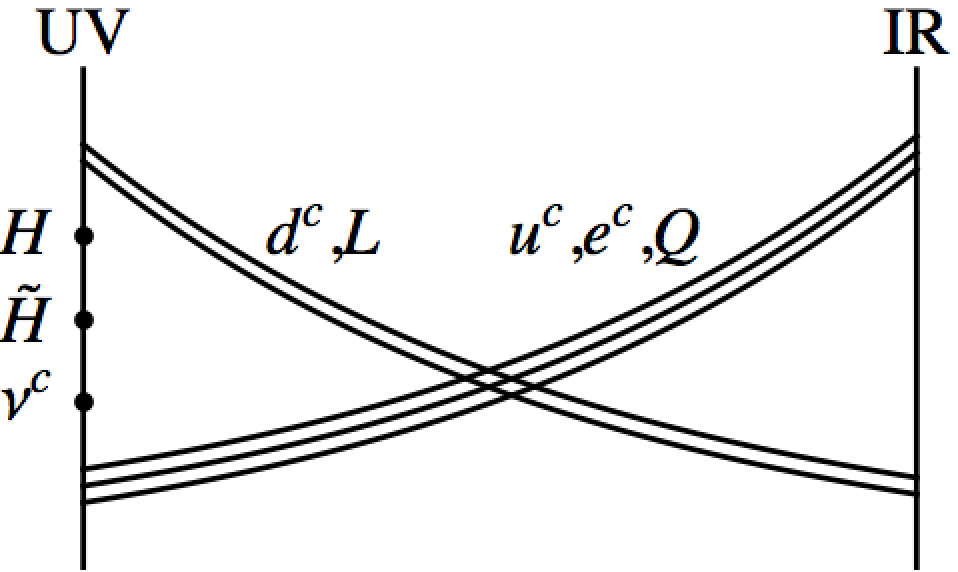}
\caption{The three throat picture.  \emph{Left:} The $Z_{3}$ flavour symmetry has a geometric description in terms of a rotational symmetry between throats.  Hierarchies and flavour symmetry breaking arise from different throat lengths, corresponding to dynamical scales in the dual CFT.  \emph{Right:} The localisation of fields in each throat.  Fields on or near the IR brane are composite operators in the dual CFT while those on or near the UV brane are elementary.\label{fig:WEDtthroat1}}
\end{center}
\end{figure}

Though there are many possibilities in this general setup, we will consider a model similar to the field theory discussed in \S\ref{sec:CH}, so as to allow for an explicit comparison that highlights the strong coupling effects in both models.  Accordingly, we choose to localise the right handed down quarks and lepton doublets ($d^c$ and $L$) near the UV brane, and the right handed up quarks, charged leptons and quark doublets ($u^c$, $e^c$ and $Q$) near the IR brane.  A single generation of Higgs fields and three generations of right handed neutrino ($H$, $\tilde{H}$ and $\nu^{c}$) are all fixed on the UV brane.  A graphical representation of the model is given in Figure \ref{fig:WEDtthroat1}.  In the dual, strongly coupled CFT this model corresponds to mainly composite fields $u^c$, $e^c$ and $Q$, and mainly elementary fields $d^c$, $L$, $\nu^c$, $H$ and $\tilde{H}$ just as in the field theory example.

In order to calculate the coupling constants in this picture we make use of the results in Refs.~\cite{Marti:2001iw, Gherghetta:2000qt, Gherghetta:2006ha} to find
\be\label{eq:WEDW4d}
W_4=W_{\rm UV}\sbrack{\Phi_4(x^{\mu}),\Phi_5(0,x^{\mu})}+\sum_{i=1}^3e^{-3kR_i}W_{\rm IR}^{(i)} \sbrack{e^{k\pi R_i}\Phi_4(x^{\mu}),\Phi_5(R_i,x^{\mu})}
\ee
where $\Phi_4$ denotes the 4D fields on the branes (which pick up normalisation factors $e^{k\pi R_i}$ from the kinetic terms on the IR branes) and $\Phi_5$ denotes the 5D fields in the bulk.  Their zero modes are given by
\be\label{eq:WED0m}
\Phi_5(y_i,x^{\mu})=Nke^{\nbrack{\frac{3}{2}-c}k\pi y_i}\Phi_4(x^{\mu})\sep{where}N=\sqrt{\frac{(1-2c)}{e^{(1-2c)k\pi R_i}-1}}\,.
\ee
Here, $N$ is a normalisation factor arising from the kinetic terms and $c$ is the bulk mass of the field; it is less than $1/2$ for IR localised fields and greater than $1/2$ for UV localised fields.  It is generally difficult to have superpotential couplings appearing in the bulk as the notion of supersymmetry in five dimensions is quite different from the four dimensional interpretation on the branes.  In our case, the superpotential is entirely localised on the UV brane:
\be
W_{\rm UV}=\mu\tilde{H}H+\lambda_{d,ij}^5d^c_{5,i}Q_{5,j}\tilde{H}+\lambda_{u,ij}^5u^c_{5,i}Q_{5,j}H+ \lambda_{e,ij}^5e^c_{5,i}L_{5,j}\tilde{H}+\lambda_{\nu,ij}^{5}\nu^c_iL_{5,j}H+M_{\nu,ij}\nu^c_i\nu^c_j
\ee
with the forms all UV coupling constants being determined by the $Z_{3}$ symmetry on the UV brane.  Plugging into Eq.~\eqref{eq:WEDW4d} we thus derive the following expressions for the 4D coupling constants
\ba
\lambda_{d,ij} &=& \lambda_{d,ij}^5kN_{d_{i}}N_{Q_{j}}\nonumber\\
\lambda_{u,ij} &=& \lambda_{u,ij}^5kN_{u_{i}}N_{Q_{j}}\nonumber\\
\lambda_{e,ij} &=& \lambda_{e,ij}^5kN_{e_{i}}N_{L_{j}}\nonumber\\
\lambda_{\nu,ij} &=& \lambda_{\nu,ij}^5kN_{L_j}
\ea
with the normalisation factors given in Eq.~\eqref{eq:WED0m}.

Since the right handed neutrinos live on the UV brane they have no knowledge of the $Z_3$ breaking in the throats and their Majorana mass is determined as in Eq.~\eqref{eq:CHxi}, with the mass scale $m$ being replaced by the UV scale.  The left handed neutrinos have a presence in the throats so their Yukawa could pick up a $Z_3$ breaking contribution through the normalisation factors $N_{L_{i}}$ (which would correspond to a hierarchy in the columns of $\lambda_{\nu}$).  However, as long as the lepton doublet is well localised in the UV (i.e.\ $c_L\gg1/2$) this contribution is negligible.  The neutrino mass matrix is therefore determined in exactly the same way as in the field theory case (see Eq.~\eqref{eq:CHmnu}).  Again, we need to force a diagonal charged lepton Yukawa but we will discuss how this can arise naturally shortly.  Hence for a UV scale of around the GUT scale we easily recover the previous results in the neutrino sector: tribimaximal mixing and eV scale light neutrino masses given by Eq.~\eqref{eq:CHmnuev}.

Meanwhile in the quark and charged lepton sectors there is a strong presence in the throats so the Yukawa couplings feel the full effects of the hierarchical warping factors and the subsequent $Z_{3}$ breaking.  For the 4D couplings to be consistent with the third generation masses we need the third generation Yukawa couplings to be of order unity, i.e.\ there cannot be much warping in the shortest throat, which contains the third generation fields.  To this end we choose $k\pi R_3=1$ in which case $N_{3}\sim1$ for all fields propagating in the third throat.  We also assume $\lambda^{5}k\sim1$ for each bulk coupling.  Assuming that the warping effect is stronger in the remaining throats, we can then approximate the Yukawa couplings as
\be
\lambda_d=
\nbrack{\begin{array}{ccc}
a\varepsilon_d\eta_d & b\varepsilon_d & c \\
c\varepsilon_d\eta_d & a\varepsilon_d & b \\
b\varepsilon_d\eta_d & c\varepsilon_d & a \end{array}},\quad
\lambda_u=\nbrack{\begin{array}{ccc}
a\varepsilon_u\eta_d & b\varepsilon_u & c \\
c\varepsilon_u\eta_d & a\varepsilon_u & b \\
b\varepsilon_u\eta_d & c\varepsilon_u & a \end{array}},\quad
\lambda_e=\nbrack{\begin{array}{ccc}
a\varepsilon_e\eta_d & b\varepsilon_e & c \\
c\varepsilon_e\eta_d & a\varepsilon_e & b \\
b\varepsilon_e\eta_d & c\varepsilon_e & a \end{array}}
\ee
where the $\varepsilon$'s and $\eta$'s break the $Z_{3}$ symmetry and arise from the warp factors:
\begin{align}
\varepsilon_d & \approx e^{-\nbrack{\frac{1}{2}-c_Q}k\pi(R_2-R_3)} & \varepsilon_u & \approx e^{-\nbrack{1-c_{u}-c_Q}k\pi(R_2-R_3)} & \varepsilon_e & \approx e^{-\nbrack{\frac{1}{2}-c_e}k\pi(R_2-R_3)} \nonumber\\
\eta_d & \approx e^{-\nbrack{\frac{1}{2}-c_Q}k\pi(R_1-R_2)} & \eta_u & \approx e^{-\nbrack{1-c_{u}-c_Q}k\pi(R_1-R_2)} & \eta_e & \approx e^{-\nbrack{\frac{1}{2}-c_e}k\pi(R_1-R_2)}\,.
\end{align}
The other parameters in the Yukawas are assumed arbitrary and of order unity.  As in the field theory, the form of the bulk couplings $\lambda^5$ is set by the $Z_{3}$ symmetry but this time one immediately sees that there are three \emph{independent} hierarchies.  However, if we assume an underlying SU(5) GUT like structure we expect to find $c_{e}=c_{u}=c_{Q}$ and $c_{d}=c_{L}$.  Substituting into the above equation immediately suggests that $\varepsilon_{d}=\varepsilon_{e}=\sqrt{\varepsilon_{u}}$ and $\eta_{d}=\eta_{e}=\sqrt{\eta_{u}}$ thus reproducing the structure found in \S\ref{sec:CH}.  As such, the same arguments can be used to show that this model does indeed reproduce the MSSM masses and mixings.

Comparing the warping factors with the measured mass values \eqref{eq:exmasses} one can get a handle on the lengths of the other throats\footnote{The existence of flavour changing neutral currents (arising from couplings to the gluon Kaluza-Klein modes) provides an upper bound on the throat length.  Ref.~\cite{Delgado:1999sv} gives the absolute worst case scenario of $1/R\gtrsim300\mbox{ TeV}$.}:
\begin{align}
\nbrack{\frac{1}{2}-c_Q}k\pi(R_2-R_3) & \sim\ln{\nbrack{\frac{m_b}{m_s}}}\sim\ln{\nbrack{\frac{m_{\tau}}{m_{\mu}}}}\sim\frac{1}{2}\ln{\nbrack{\frac{m_t}{m_c}}} \nonumber\\
\nbrack{\frac{1}{2}-c_Q}k\pi(R_1-R_2) & \sim\ln{\nbrack{\frac{m_s}{m_d}}}\sim\ln{\nbrack{\frac{m_{\mu}}{m_{e}}}}\sim\frac{1}{2}\ln{\nbrack{\frac{m_{c}}{m_{u}}}}
\end{align}
and therefore
\be
\frac{R_1-R_2}{R_2-R_3}\sim0.99\sim1.31\sim1.9\,.
\ee
These relations are not exact, but it should be noted that we are considering a simplified picture.  In reality there are several other effects that would effect the flavour structure (such as tunnelling between throats \cite{Dimopoulos:2001ui, Dimopoulos:2001qd} and back reactions from the spontaneous $Z_{3}$ breaking) that contribute to the Yukawa couplings.  It is entirely conceivable that such effects will improve these results by the required factors of order unity. 

\begin{figure}[!t]
\begin{center}
\includegraphics[width=4cm]{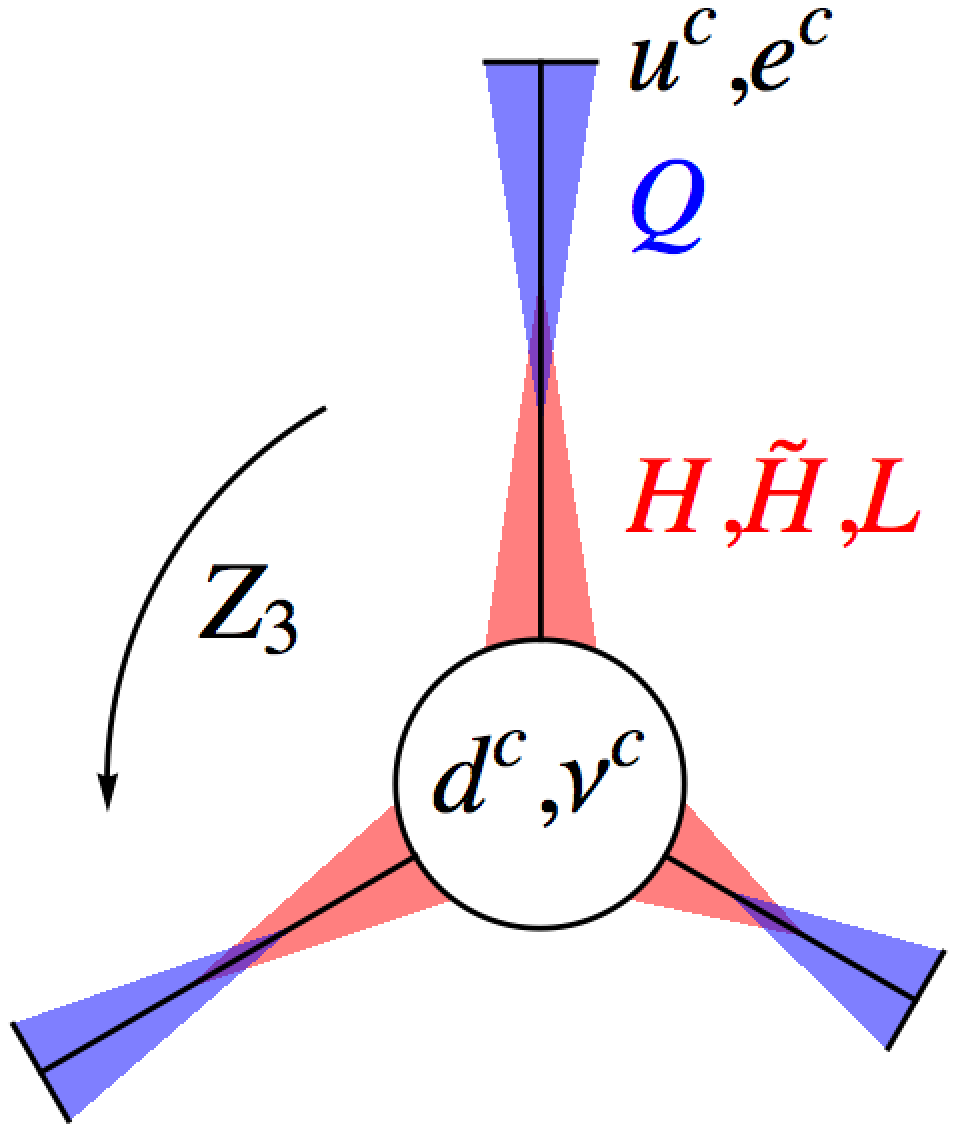}\hspace{2cm}
\includegraphics[width=6cm]{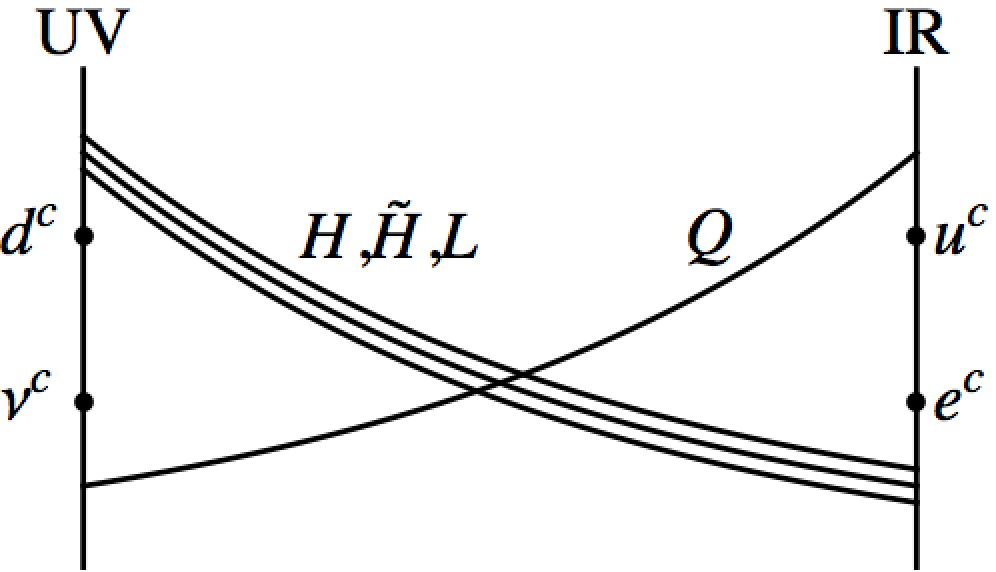}
\caption{The three throat picture with diagonal charged lepton Yukawa.  The charged lepton Yukawa lives at the end of each throat, so only couplings to the fields in the same throat are allowed.\label{fig:WEDtthroat2}}
\end{center}
\end{figure}

As mentioned earlier, by implementing an alternative approach one can arrange for the charged lepton Yukawa to be naturally diagonal.  The price for this convenience is a loosening of the precise SU(5) GUT structure we found in the previous example.  Consider instead the scenario represented in Figure \ref{fig:WEDtthroat2}.  Here each generation of right handed up quarks and charged leptons are confined to the IR branes at the end of their throats.  The only fields that can couple to them are those whose wave function has an overlap with this IR brane, i.e.\ those which propagate in the same throat.  As such, both the up quark and charged lepton Yukawa are necessarily diagonal in the throat basis.  Conversely, the right handed down quarks and neutrinos are confined to the UV brane.  All three throats have an intersection here so different generations are allowed to mix.  Thus all mixing in this (and any other) multiple throat model occurs \emph{only} on the UV brane.

For this picture to work out we must allow a single generation of Higgs fields to propagate in all three throats as described in Ref.~\cite{Cacciapaglia:2006tg} but this simply results in a normalisation factor
\be
N=\sqrt{\frac{(1-2c)}{e^{(1-2c)k\pi R_1}+e^{(1-2c)k\pi R_2}+e^{(1-2c)k\pi R_3}-3}}
\ee
for the Higgs fields.  The superpotential now has both an IR and UV component so is a little more complicated, but using the results in this section one can easily show that
\be
\lambda_d=
\nbrack{\begin{array}{ccc}
a\varepsilon_d\eta_d & b\varepsilon_d & c \\
c\varepsilon_d\eta_d & a\varepsilon_d & b \\
b\varepsilon_d\eta_d & c\varepsilon_d & a \end{array}},\quad
\lambda_u=x_{u}\nbrack{\begin{array}{ccc}
\varepsilon_u\eta_u & 0 & 0 \\
0 & \varepsilon_u & 0 \\
0 & 0 & 1 \end{array}},\quad
\lambda_e=a_{e}\nbrack{\begin{array}{ccc}
\varepsilon_e\eta_e & 0 & 0 \\
0 & \varepsilon_e & 0 \\
0 & 0 & 1 \end{array}}
\ee
where the $\varepsilon$'s and $\eta$'s again break the $Z_{3}$ symmetry and arise from the warp factors:
\begin{align}
\varepsilon_d & \approx e^{-\nbrack{\frac{1}{2}-c_Q}k\pi(R_2-R_3)} & \varepsilon_u & \approx e^{-\nbrack{c_H-\frac{1}{2}}k\pi(R_2-R_3)} & \varepsilon_e & \approx e^{-\nbrack{c_L+c_{\tilde{H}}-1}k\pi(R_2-R_3)} \nonumber\\
\eta_d & \approx e^{-\nbrack{\frac{1}{2}-c_Q}k\pi(R_1-R_2)} & \eta_u & \approx e^{-\nbrack{c_H-\frac{1}{2}}k\pi(R_1-R_2)} & \eta_e & \approx e^{-\nbrack{c_L+c_{\tilde{H}}-1}k\pi(R_1-R_2)}\,.
\end{align}
Since there is in principle no GUT symmetry to constrain the $c$'s in this case the warp factors remain independent and the various MSSM hierarchies are easily reproduced.
However it should be noted that the {\em approximate} underlying GUT structure is retained in the manner described in the Introduction: all the fields that appear within the same GUT multiplet have their wave functions localised near each other.

\begin{figure}[!t]
\begin{center}
\includegraphics[width=5.5cm]{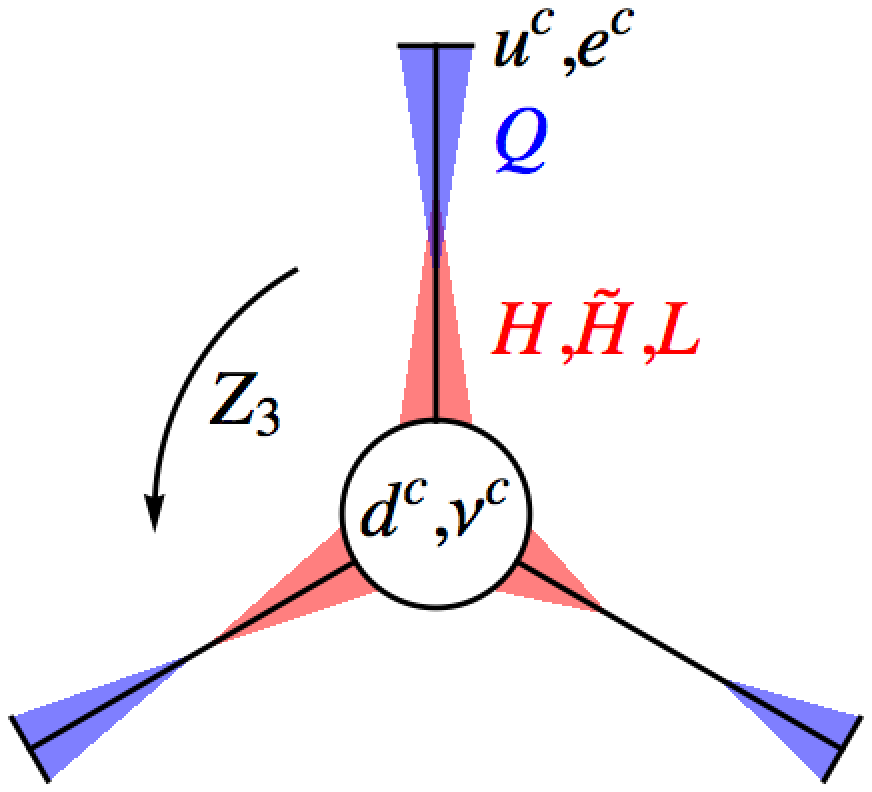}
\caption{An alternative three throat picture.  Hierarchies and flavour symmetry breaking arise from different field localisations instead of different throat lengths.  This corresponds to varying the dimensions of operators in the CFT.\label{fig:WEDcthroat}}
\end{center}
\end{figure}

Note also that one can create hierarchies by using the bulk masses rather than the throat lengths.  Since it is the product $ck\pi R$ that appears in the exponent of the warp factors, we could allow different bulk masses $c$ in each throat and keep $R$ fixed to arrive at exactly the same result.  In this scenario having multiple throats is not strictly necessary but continues to give a clear, geometrical interpretation of the flavour symmetry and where it is broken.  However, the strong coupling interpretation is now quite different and more akin to \S\ref{sec:AH}.  The bulk mass is related to the dimension of the scalar component of the corresponding CFT operator via
\be
{\rm dim}\mathcal{O}=\frac{3}{2}-c\,.
\ee
Hence small values of $c$ (i.e.\ IR brane localisation) in the AdS theory yield higher dimensional operators in the CFT.  Consequently any superpotential interactions between these fields and elementary fields will be more suppressed for smaller values of $c$.  This can come about in more than one way.  For example, Refs.~\cite{Kaplan:1997tu, Haba:1997bj, Haba:1998wf} explicitly build Standard Model fields out of operators of increasing dimension.  Alternatively the effect could be tied to the strong coupling in a highly non-trivial way such as through the anomalous dimensions of the operators (see \S\ref{sec:AH} or Ref.~\cite{Nelson:2000sn} for example).  Regardless, such a situation is well described in the AdS picture and thus can be easily represented in the same, diagrammatic fashion.  See Figure \ref{fig:WEDcthroat} for an example.

Finally it is worth mentioning how this picture applies to the non-supersymmetric case.  All arguments regarding the flavour structure are directly transferable.  In fact we have more model building freedom as there is no need to confine couplings to either the UV or IR branes.  For example, in the non-supersymmetric case we could take the model represented in Figure \ref{fig:WEDtthroat2} and move the right handed quarks and charged leptons slightly into the bulk to join the left handed quark doublets.  At the same time we could move the right handed down quarks slightly into the bulk to join the lepton doublets.  This would combine the GUT structure of our first example with the naturally diagonal charged lepton Yukawa of our second to create a very satisfying model.  The problem lies in the Higgs sector.  Since there is always a short throat with an order unity warping factor it is impossible to use the warped extra dimensions to stabilise the Higgs mass.  Hence supersymmetry, or some other mechanism, will generically be required to solve the Higgs mass hierarchy problem in these models.

\section{Conclusions}

In this paper we have demonstrated how the two principles of strong coupling and discrete symmetry can be used to explain the flavour structure observed in the Standard Model.  All hierarchies, both in particle masses and the quark mixing matrix, arise from strong coupling effects.  Tribimaximal mixing occurs in the neutrino sector thanks to a discrete flavour symmetry between generations.  This flavour symmetry is respected by the \emph{whole model} at high scales, however it is broken by the strongly coupled physics in the low energy theory.  By building models where the quarks and charged leptons feel the strong coupling but the neutrinos do not, it is thus possible to simultaneously find tribimaximal neutrino mixing \emph{and} a realistic CKM matrix using only a single flavour symmetry.  Such models are readily compatible with grand unified theories.

We provided two explicit examples, using $\cN=1$ supersymmetry and $s$-confinement to understand the strongly coupled physics.  The first (see \S\ref{sec:CH}) attaches a distinct strong coupling sector to each generation -- hierarchies are generated by the three different dynamical scales.  The second (see \S\ref{sec:AH}) assigns a different anomalous dimension to each generation to arrive at analogous hierarchies.  In both cases the hierarchies are natural and generated using only parameters of order unity.  By including a discrete $Z_{3}$ permutation symmetry between generations tribimaximal mixing occurs in the neutrino sector, where the strong coupling has no influence.  Meanwhile in the quark and charged lepton sectors strong coupling breaks the flavour symmetry and we find realistic values for the particle masses and CKM matrix.  If we assume the up quark Yukawas come from composite-composite-elementary operators in the UV while the down quark and charged lepton Yukawas come from composite-elementary-elementary operators (as would necessarily be the case in the standard SU(5) GUT), we naturally expect the the mass hierarchies in the down quark and charged leptons sectors to be similar and roughly the square root of the up quark mass hierarchy.  Both of these models predict two degenerate light neutrino masses at tree level, in keeping with the experimental observation $\Delta m_{21}^2\ll\Delta m_{32}^2$.

\begin{figure}[!th]
\begin{center}
\includegraphics[width=4cm]{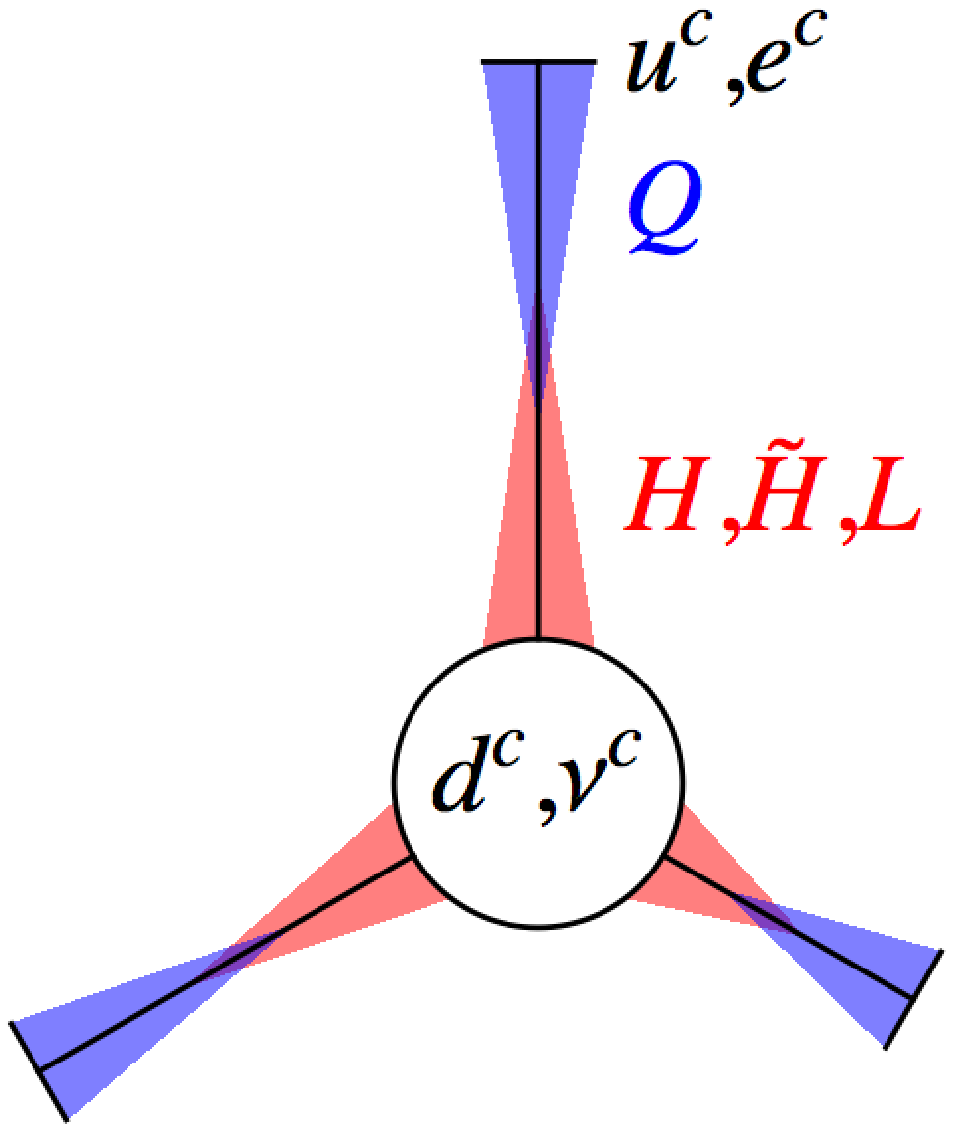}
\includegraphics[width=4cm]{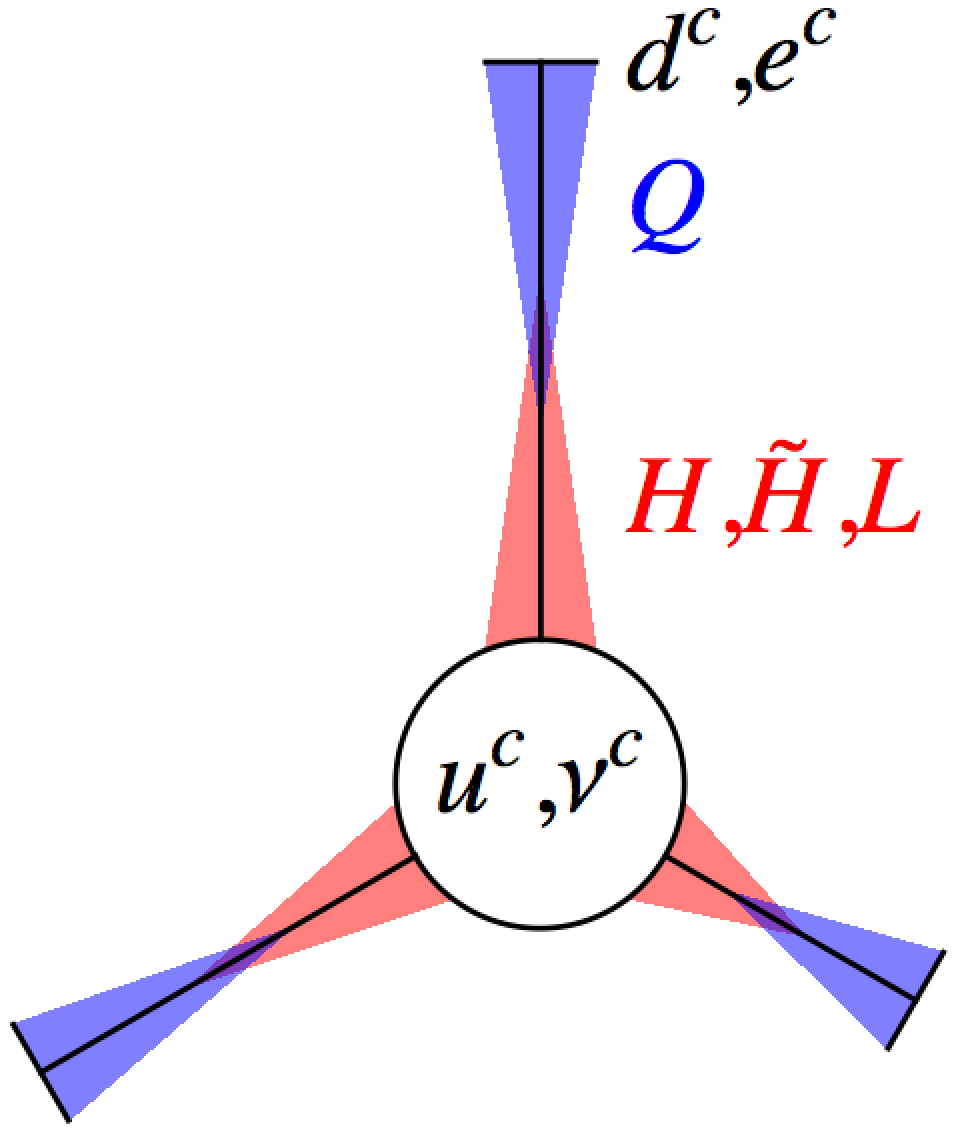}
\includegraphics[width=4cm]{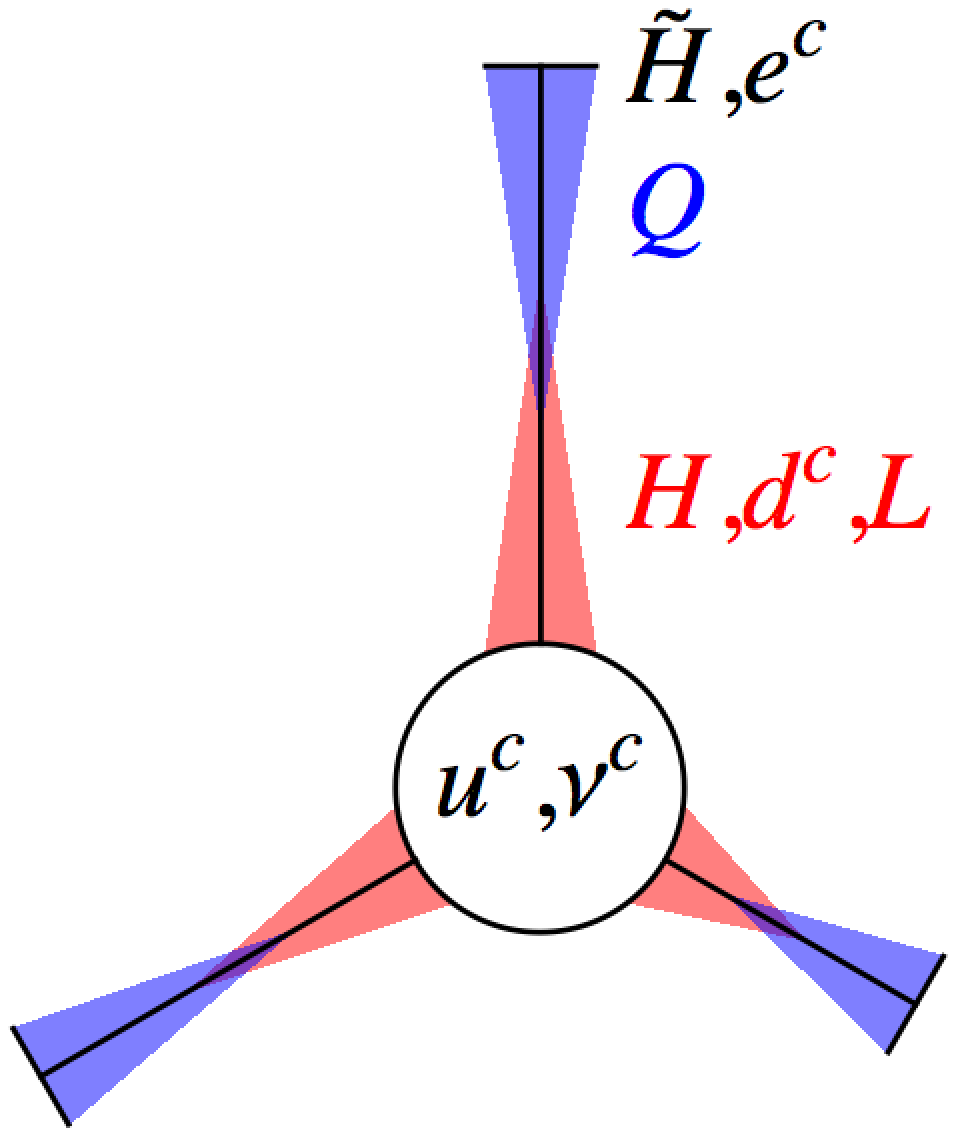}
\caption{The three throat picture can be realised with many different matter distributions.\label{fig:Ctthroat3}}
\end{center}
\end{figure}

Equivalent geometrical models can be built by appealing to the AdS/CFT correspondence.  The appropriate spacetime background in the 5D theory consists of a central UV brane from which several warped throats emanate.  In our case there is a network of three such throats -- one for each generation.  The $Z_{3}$ flavour symmetry then has a geometrical interpretation as a rotational symmetry between throats, whereas hierarchies are represented by different throat lengths or different localisations of fields within the throats: both options being understood as a spontaneous breaking of the discrete symmetry.  Only the fields localised in the throats are affected so, if the neutrinos are all localised on the central UV brane, tribimaximal mixing is maintained.  On the CFT side the throat lengths and field localisations correspond to dynamical scales and anomalous dimensions respectively, hence this picture is totally analogous to (and gives a geometric description of) the $s$-confinement models.

There are many model building possibilities in this general setup and all can be represented graphically (see Figure \ref{fig:Ctthroat3} for several $Z_{3}$-based examples that reproduce the correct flavour structure).  This provides a powerful tool.  For example, by varying the geometry of the throat network one could quickly construct theories with alternative discrete flavour symmetry groups -- such as $A_{4}$ or $S_{4}$.

\subsection*{Acknowledgements}

The authors would like to thank Tony Gherghetta and Michael Schmidt for discussions. SAA is supported by a Leverhulme research fellowship. 

\appendix

\section{$Z_3$ and the UV superpotential\label{app:Z3}}

To derive the expressions \eqref{eq:CHxi} for the coupling constants appearing in the UV superpotential we first review a few aspects of the discrete $Z_3$ permutation symmetry.  Acting on the triplet $(Y_1,\,Y_2,\,Y_3)$ the group performs cyclic permutations and can be represented by the action of the matrix
\be
T=\nbrack{\begin{array}{ccc}
0 & 1 & 0 \\
0 & 0 & 1 \\
1 & 0 & 0 \end{array}}.
\ee
To respect the $Z_3$ symmetry of the model all terms in the superpotential \eqref{eq:CHWuv} must be invariant under such transformations.  The affected terms are
\ba
W_{\rm UV} &\supset& \xi_{P,ia}\tilde{P}_iY_aZ_a+\frac{1}{M_X}\xi_{Q,ia}\tilde{Q}_iY_aY_a\tilde{H}+\frac{1}{M_X^2}\zeta_{ab}\epsilon^{(5)}Y_aY_aY_bY_bH+\nonumber\\
&& \lambda_{\nu,ij}\nu^c_i\tilde{Q}_jH+M_{\nu,ij}\nu^c_i\nu^c_j\,.
\ea
First consider the term $\xi_Q\tilde{Q}YY\tilde{H}$ that contains the down quark and charged lepton Yukawas.  The $Z_3$ structure of this term is
\be
(\tilde{Q}_1,\,\tilde{Q}_2,\,\tilde{Q}_3)\xi_Q
\nbrack{\begin{array}{c} Y_1Y_1 \\ Y_2Y_2 \\ Y_3Y_3 \end{array}}.
\ee
Since the $Y$'s couple diagonally $T$ permutes the triplet $(Y_1Y_1,\,Y_2Y_2,\,Y_3Y_3)$ exactly as it would the original triplet $(Y_1,\,Y_2,\,Y_3)$.  To ensure the term is invariant we thus require
\be\label{eq:CHxiQ}
T\xi_QT^T=\xi_Q\quad\implies\quad
\xi=\nbrack{\begin{array}{ccc}
a & b & c \\
c & a & b \\
b & c & a \end{array}}
\ee
for some undetermined arbitrary constants $a$, $b$ and $c$.  Exactly the same argument holds for $\lambda_{\nu}$ and $\xi_P$.  For the remaining matrices $\zeta$ and $M_{\nu}$ there is an additional constraint arising from the fact that the matrices must be symmetric.  This is clear in case of $M_{\nu}$, but for $\zeta$ one must notice that the contraction of the gauge indices with an alternating tensor results in the term $\zeta YYYYH$ being symmetric in the indices $a$ and $b$.

\section{Deviations from tribimaximal mixing\label{app:dtbm}}

In \S\ref{sec:CHmass} we considered the simplest case in the lepton sector, where $b=c=0$ in the charged lepton Yukawa coupling of Eq.~\eqref{eq:CHlambda}.  As it stands there is no symmetry principle enforcing this condition (although such a choice could be justified by, for example, embedding the $Z_3$ symmetry in and SU(3) flavour symmetry or through the geometric arguments of \S\ref{sec:WED}) so one might like to consider deviations from equality.  Suppose we allow off diagonal entries of order $\sigma$.  The charged lepton mass matrix is modified to
\be
m_l\sim\frac{\expect{\tilde{H}}\Lambda_3}{M_X}\nbrack{\begin{array}{ccc}
a_e\varepsilon\eta & \sigma\varepsilon & \sigma \\
\sigma\varepsilon\eta & a_e\varepsilon & \sigma \\
\sigma\varepsilon\eta & \sigma\varepsilon & a_e \end{array}}.
\ee
In general, the new charged lepton mass matrix will require a bi-unitary transformation to put in into a diagonal form in which case the PMNS matrix will be adjusted accordingly.  The relevant unitary matrix is the one acts on the left handed lepton doublet, i.e.\ the one that diagonalises
\be
m_lm_l^{\dag}\sim\nbrack{\frac{\expect{\tilde{H}}\Lambda_3}{M_X}}^2\nbrack{\begin{array}{rrr}
\varepsilon^2\eta^2+\sigma^2 & \sigma\varepsilon^2+\sigma^2 & \sigma \\
\sigma\varepsilon^2+\sigma^2 & \varepsilon^2+\sigma^2 & \sigma \\
\sigma & \sigma & 1 \end{array}}
\ee
where we have assumed that $a_e\sim1$ and that $\sigma,\varepsilon,\eta\ll1$.  We thus find
\be
U_l\sim\nbrack{\begin{array}{ccc}
1 & \frac{\sigma\varepsilon^2+\sigma^2}{\varepsilon^2+\sigma^2} & \sigma \\
\frac{\sigma\varepsilon^2+\sigma^2}{\varepsilon^2+\sigma^2} & 1 & \sigma \\
\sigma & \sigma & 1 \end{array}}.
\ee
If $\sigma<\varepsilon$ the $1,2$ component goes like $\sigma$, but if $\sigma<\varepsilon$ it is of order unity.  Assuming $\sigma<\varepsilon$, the net result is to change the unitary transformation required to diagonalise the neutrino mass matrix.  Eq.~\eqref{eq:CHmnud} becomes
\be
\hat{m}_{\nu}=U_{\nu}^{\dag}U_l^{\dag}m_{\nu}U_lU_{\nu}.
\ee
We know from \S\ref{sec:CHmass} that the matrix $U_lU_{\nu}$ is equal to the tribimaximal mixing matrix $V_{\rm HPS}$ given in Eq.~\eqref{eq:PMNSex}.  In other words, the neutrino mixing matrix becomes
\be
V_{\rm PMNS}=U_l^{\dag}V_{\rm HPS}\sim V_{\rm HPS}+\mathcal{O}(\sigma)\,.
\ee
Since $U_l$ is a unitary transformation, the eigenvalues of the neutrino mass matrix are unchanged.

\section{CP violation\label{app:J}}

It is convenient to parameterise the CP violation using the Jarlskog invariant \cite{Jarlskog:1985cw, Amsler:2008zzb}, which is given by
\be\label{eq:Jarl}
J=\frac{i\det{[M_u^{\dag}M_u,M_d^{\dag}M_d]}}{2F_uF_d}=\nbrack{3.05^{+0.19}_{-0.20}}\times10^{-5}
\ee
and is non-zero if and only if CP violation is present in the quark sector.  The matrices $M_{u/d}$ are the quark mass matrices normalised by the top and bottom quark masses, whereas the quantites $F_u$ and $F_d$ are defined to be
\be
F_{u/d}=\sbrack{1-\nbrack{\frac{m_{c/s}}{m_{t/b}}}^2} \sbrack{1-\nbrack{\frac{m_{u/d}}{m_{t/b}}}^2} \sbrack{\nbrack{\frac{m_{c/s}}{m_{t/b}}}^2-\nbrack{\frac{m_{u/d}}{m_{t/b}}}^2}\,.
\ee

\bibliographystyle{JHEP-2}
\bibliography{../masterbib}
\end{document}